\documentclass[12pt]{article}

\title{Relaxation of the distribution function tails for systems described by Fokker-Planck equations}

\usepackage{amsthm,amsmath,amssymb,a4wide,psfig,epsf}
\def\mb#1{\setbox0=\hbox{$#1$}\kern-.025em\copy0\kern-\wd0
\kern-0.05em\copy0\kern-\wd0\kern-.025em\raise.0233em\box0}

\def\ds{\displaystyle}

\begin{document}

\author{Pierre-Henri Chavanis$^{1}$ and Mohammed Lemou$^{2}$}
\maketitle
\begin{center}
(1) Laboratoire de Physique Th\'eorique (UMR 5152 du CNRS), Universit\'e
Paul Sabatier,\\ 118, route de Narbonne, 31062 Toulouse Cedex 4, France\\
E-mail: {\it chavanis{@}irsamc.ups-tlse.fr} \\
(2) Math\'ematiques pour l'Industrie et la Physique (CNRS UMR~5640),\\
Universit\'e Paul Sabatier
118, route de Narbonne,\\
31062 Toulouse Cedex 4, France\\
E-mail: {\it lemou{@}mip.ups-tlse.fr}

\vspace{0.5cm}
\end{center}

\begin{abstract}

We study the formation and the evolution of velocity distribution
tails for systems with weak long-range interactions. In the thermal
bath approximation, the evolution of the distribution function of a
test particle is governed by a
Fokker-Planck equation where the diffusion coefficient depends on the
velocity. We extend the theory of Potapenko {\it et al.} 
[Phys. Rev. E, {\bf 56}, 7159 (1997)] developed for power-law
diffusion coefficients to the case of an arbitrary form of diffusion
coefficient and friction force. We study how the structure and the
progression of the front depend on the behavior of the diffusion
coefficient and friction force for large velocities. Particular
emphasis is given to the case where the velocity dependence of the
diffusion coefficient is Gaussian. This situation arises in
Fokker-Planck equations associated with one dimensional systems with
long-range interactions such as the Hamiltonian Mean Field (HMF) model
and in the kinetic theory of two-dimensional point vortices in
hydrodynamics. We show that the progression of the front is extremely
slow (logarithmic) in that case so that the convergence towards the
equilibrium state is peculiar. Our general formalism can have applications
for other physical systems such as optical lattices.

\end{abstract}

\section{Introduction}
\label{sec_introduction}

The study of Fokker-Planck equations is an important problem in
statistical mechanics and kinetic theory \cite{risken}. In the
simplest models, the diffusion coefficient is constant. However,
Fokker-Planck equations with a diffusion coefficient depending on the
velocity of the particles have also been introduced in physics.  These
equations usually describe the relaxation of a ``test particle''
evolving in a bath of ``field particles'' at statistical equilibrium
when the particles interact via weak long-range forces. In that case,
the diffusion coefficient is a function of the velocity of the test
particle.  For example, in his Brownian theory of stellar dynamics,
Chandrasekhar \cite{chandra} describes the evolution of the velocity
distribution of a star in a cluster by a Fokker-Planck equation
involving a diffusion and a friction. The coefficients of diffusion
and friction are related to each other by an Einstein relation and the
diffusion coefficient decreases as $v^{-3}$ for large
velocities. These results are similar to those obtained in plasma
physics for the Coulombian interaction \cite{balescu}. By using an
analogy with stellar dynamics, Chavanis
\cite{drift,kin,houches} describes the relaxation of a test vortex
in a thermal bath of field vortices by a Fokker-Planck equation (in
position space) involving a diffusion and a drift along the vorticity
gradient.  The coefficients of drift and diffusion are related to each
other by a form of Einstein relation involving a negative temperature
and the diffusion coefficient is inversely proportional to the local
shear created by the vortex cloud. For a Gaussian distribution of
field vortices, the diffusion coefficient of the test vortex decreases
with the distance as $r^{2} e^{-\lambda r^{2}}$
\cite{hb,kinlemou}. Similarly, for the Hamiltonian Mean Field (HMF) 
model, Bouchet \& Dauxois \cite{bouchet,bd} and Chavanis {\it et al.} 
\cite{cvb} find that the velocity distribution of a test particle
satisfies a Fokker-Planck equation with a diffusion coefficient
decreasing as $e^{-\beta v^{2}/2}$ for large velocities. More
generally, using the theory developed by Landau, Lenard and Balescu in
plasma physics
\cite{ichimaru} and implementing a thermal bath approximation, one can
obtain a general Fokker-Planck equation involving an anisotropic
diffusion coefficient depending on the velocity of the test
particle. This Fokker-Planck equation is valid for systems with weak
long-range potentials of interaction. The preceding kinetic equations
can be recovered as particular cases of this general Fokker-Planck
equation \cite{hb}.

For Fokker-Planck equations with a variable diffusion coefficient, the
relaxation towards the Boltzmann distribution is slowed down,
especially if the diffusion coefficient decreases rapidly with the
velocity. One consequence is that velocity correlation functions can
decrease {\it algebraically} rapidly with time (instead of
exponentially) as investigated by Bouchet \& Dauxois \cite{bd} in
relation with the HMF model \footnote{In that case, the Fokker-Planck
equation with Gaussian diffusion coefficient and linear friction can
be transformed into a Fokker-Planck equation with constant diffusion
coefficient and logarithmic potential which is known to exhibit
power-law correlations (see, in particular, Appendix B of \cite{mark} and
\cite{lutz}).}. They therefore explain the observed algebraic
tails of the velocity correlations functions in terms of classical
kinetic theory without invocating a notion of ``generalized
thermodynamics''. Here, we consider the relaxation of the system
towards equilibrium from another point of view. We focus on the
distribution function $f(v,t)$ and study the structure and the
evolution of the front formed in the high velocity tail. Our study is
based on the approach of Potapenko {\it et al.} \cite{potapenko} who
studied this problem in the case where the diffusion coefficient
decreases algebraically with the velocity. In the present paper, we
generalize this approach to an arbitrary form of diffusion coefficient
and study how the front position $v_{f}(t)$ and the front structure
evolve with time. In the case of a Gaussian or exponential decay of
the diffusion coefficient with the velocity, we find that the
progression of the front is extremely slow (logarithmic).  This can
leads to a sort of ``kinetic blocking'' or, at least, a ``slowing
down'' of the relaxation. A similar confining effect was noted in the
case where the diffusion coefficient depends on the density \cite{csr,rr}, 
but this situation (related to the theory of violent relaxation) 
is more difficult to analyze since the kinetic equation is then nonlinear.

The paper is organized as follows. In Sec. \ref{sec_ex}, we discuss
different systems with long-range interactions (stellar systems,
Coulombian plasmas, two-dimensional vortices, HMF model,...) which are
described in the thermal bath approximation by Fokker-Planck equations
with a variable diffusion coefficient. We also consider systems like
optical lattices described by Fokker-Planck equations with constant
diffusion coefficient but variable friction coefficient. In Sec.
\ref{sec_gen}, we generalize the theory of Potapenko {\it et al.}
\cite{potapenko} for an arbitrary form of diffusion coefficient 
and friction force. We provide general equations characterizing the
structure and the evolution of the front for large times. In
Sec. \ref{sec_val}, we address the validity of our approach. In
Sec. \ref{sec_part}, we consider particular applications of our
general formalism for physically motivated Fokker-Planck
equations. Analytical results are compared with direct numerical
simulations of the Fokker-Planck equation.  Particular emphasis is
given to the case where the diffusion coefficient decreases with the
velocity like an exponential or a Gaussian distribution. This is the
situation relevant for one-dimensional systems like the HMF model and
for two-dimensional point vortices. Finally, in Sec. \ref{sec_exact}
we investigate a class of Fokker-Planck equations for which our
approach is exact for all times. The Appendices provide technical
details and extensions of our main results.

\section{Examples of Fokker-Planck equations with a variable
diffusion coefficient and friction force}
\label{sec_ex}

We consider a Hamiltonian system of $N$ particles interacting via a
weak long-range binary potential $u(|{\bf r}-{\bf r}'|)$. These
particles can be stars in stellar clusters, electrons or ions in a
plasma, point vortices in two-dimensional hydrodynamics, particles
located on a ring in the HMF model ... We assume that the cluster is
homogeneous and in a steady state characterized by a distribution
function $f_{0}({\bf v})$. In general, $f_{0}$ will be the statistical
equilibrium state (thermal bath) but in certain cases it can be a
slowly evolving distribution function. We introduce a ``test
particle'' and denote by $P({\bf v},t)$ its velocity distribution
function.  Due to the interaction with the ``field particles'', the
velocity distribution will change and the test particle will acquire
the distribution of the bath $f_{0}({\bf v})$ for $t\rightarrow
+\infty$ (see \cite{hb} for more details). The general Fokker-Planck
equation describing the relaxation (``thermalization'') of the test
particle in the bath of field particles can be written as \cite{hb}:
\begin{eqnarray}
\label{ex1} {\partial P\over\partial t}=\pi (2\pi)^{d}
m{\partial\over\partial v^{\mu}}\int d{\bf v}_{1}d{\bf
k}k^{\mu}k^{\nu}{\hat{u}({\bf k})^{2}\over |\epsilon({\bf k},{\bf
k}\cdot {\bf v})|^2}\delta\lbrack {\bf k}\cdot ({\bf v}-{\bf
v}_{1})\rbrack \biggl ({\partial \over\partial v^{\nu}}-{\partial
\over\partial {v}_{1}^{\nu}}\biggr )f_{0}({\bf v}_{1})P({\bf
v},t),\nonumber\\
\end{eqnarray}
where $\hat{u}({\bf k})$ is the Fourier transform of $u({\bf r})$
and
\begin{eqnarray}
\label{ex1bis} \epsilon({\bf k},\omega)=1+(2\pi)^d\hat{u}({\bf
k})\int{{\bf k}\cdot {\partial f_0\over\partial{\bf v}}\over
\omega-{\bf k}\cdot{\bf v}}d{\bf v},
\end{eqnarray}
is the dielectric function. This equation can be obtained from the
Lenard-Balescu equation by {\it fixing} the distribution function of
the field particles $f({\bf v}_{1},t)$ to its static value $f_0({\bf
v}_{1})$ \cite{balescu,hb}. This (thermal) bath approximation transforms
an integro-differential equation (Lenard-Balescu) into a differential
equation (Fokker-Planck). The Lenard-Balescu equation was introduced
in plasma physics for the Coulombian potential but it can apply to
other systems with weak long-range interactions. Equations
(\ref{ex1})-({\ref{ex1bis}) can also be obtained from the general
expression of the Fokker-Planck equation by explicitly calculating the
coefficients of friction and diffusion (first and second moments of
the velocity increments) using the Klimontovich approach
\cite{ichimaru}. The fact that
Eq. (\ref{ex1}) is linear does not imply that the distribution
$P({\bf v},t)$ is close to equilibrium. The test particle approach is
different from considering a small perturbation of the Lenard-Balescu
equation around equilibrium. In the first case, we describe the
evolution of a single test particle (or an ensemble of test particles
that do not interact among themselves) in a thermal bath while in the
second case one would describe the evolution of all the particles
(the system ``as a whole'') close to equilibrium.

If we consider that the field particles are at statistical
equilibrium with the Boltzmann distribution $f_{0}({\bf v})\sim
e^{-\beta m {v^{2}/ 2}}$, and if we neglect collective effects
taking $|\epsilon({\bf k},{\bf k}\cdot {\bf v})|=1$ (Landau
approximation) we can rewrite the general Fokker-Planck equation
(\ref{ex1}) in the form \cite{hb}:
\begin{equation}
\label{ex2} {\partial P\over\partial t}={1\over t_{R}}{\partial\over\partial x^{\mu}}\biggl\lbrack G^{\mu\nu}({\bf x})\biggl ({\partial P\over\partial x^{\nu}}+2 P x^{\nu}\biggr )\biggr\rbrack,
\end{equation}
where we have set ${\bf x}=(\beta m/2)^{1/2}{\bf v}$. The diffusion coefficient is proportional to the tensor
\begin{equation}
\label{ex3} G^{\mu\nu}({\bf x})=\int d\hat{\bf
  k}{\hat{k}^{\mu}\hat{k}^{\nu}e^{-(\hat{\bf k}\cdot {\bf
      x})^{2}}} 
\end{equation}
with $\hat{\bf k}={\bf k}/k$, and the quantity
\begin{equation}
\label{ex4} t_{R}^{-1}=\bigl ({\pi\over 8}\bigr )^{1/2}d^{3/2}(2\pi)^{d}{\rho m\over v_{m}^{3}}\int_{0}^{+\infty}k^{d}\hat{u}(k)^{2}dk
\end{equation}
provides an estimate of the inverse relaxation time of the test
particle toward the distribution of the bath. Here $v_{m}=(d/\beta
m)^{1/2}$ denotes the r.m.s. velocity and $\rho$ the spatial
density. A more general expression of the diffusion coefficient can be
obtained by taking into account collective effects \cite{hb}. Note,
however, that for $v\rightarrow
\infty$ (which is a limit that we shall be particularly interested
with in the sequel), the expression (\ref{ex3}) is asymptotically
exact since $|\epsilon({\bf k},{\bf k}\cdot {\bf v})|\rightarrow 1$
for $|{\bf v}|\rightarrow +\infty$. Note also that for weak long-range
potentials of interaction for which our approach is valid, and in the
Landau approximation, the precise form of the potential $\hat{u}(k)$
only determines the timescale of the relaxation, through
Eq. (\ref{ex4}), not the form of the kinetic operator. Therefore, in
the Landau approximation, the expression of the diffusion tensor as a
function of the velocity only depends on the dimension of space $d$.

In dimension $d=3$, the
diffusion tensor can be written
\begin{eqnarray}
G^{\mu\nu}=(G_{\|}-{1\over 2}G_{\perp}){x^{\mu}x^{\nu}\over x^{2}}+{1\over 2}G_{\perp}\delta^{\mu\nu},
\label{ex5}
\end{eqnarray}
where $G_{\|}$ and $G_{\perp}$ are the diffusion coefficients in the 
directions parallel and perpendicular to the velocity of the test particle. They are explicitly given by
\begin{eqnarray}
G_{\|}={2\pi^{3/2}\over x}G(x),\qquad
G_{\perp}={2\pi^{3/2}\over x}\lbrack {\rm erf}(x)-G(x)\rbrack,
\label{ex6}
\end{eqnarray}
with
\begin{eqnarray}
G(x)={2\over\sqrt{\pi}}{1\over x^{2}}\int_{0}^{x}t^{2}e^{-t^{2}}dt={1\over 2x^{2}}\biggl\lbrack {\rm erf}(x)-{2x\over \sqrt{\pi}}e^{-x^{2}}\biggr\rbrack,
\label{ex7}
\end{eqnarray}
where
\begin{eqnarray}
{\rm erf}(x)={2\over \sqrt{\pi}}\int_{0}^{x}e^{-t^{2}}dt,
\label{ex8}
\end{eqnarray}
is the error function.  If we consider spherically symmetric
distributions, noting that $\partial P/\partial x^{\mu}=(1/x)(\partial
P/\partial x)x^{\mu}$ and $G^{\mu\nu}x^{\nu}=G_{\|}x^{\mu}$ we obtain
\begin{equation}
\label{ex9} {\partial P\over\partial t}={1\over t_{R}}{1\over x^{2}}{\partial\over\partial x}\biggl\lbrack x^{2} G_{\|}({x})\biggl ({\partial P\over\partial x}+2 P x\biggr )\biggr\rbrack.
\end{equation}
For the gravitational potential, this Fokker-Planck equation has been
studied by Chandrasekhar in his Brownian theory of stellar dynamics
\cite{chandra}. It has also been considered in plasma physics as an
approximation of the Landau equation valid for sufficiently large
times
\cite{balescu,potapenko}. We note in particular that the diffusion coefficient
$G_{\|}(x)$ decreases algebraically like $x^{-3}$ for $x\rightarrow
+\infty$.

Alternatively, if we consider one dimensional systems ($d=1$), the
general Fokker-Planck equation (\ref{ex1}) simplifies into
\cite{hb}:
\begin{equation}
\label{ex10}  {\partial P\over\partial t}={\partial\over\partial
  v}\biggl\lbrack D(v)\biggl ({\partial P\over\partial v}-P {d\over dv}\ln f_{0}\biggr )\biggr\rbrack,
\end{equation}
where $D(v)$ is given by
\begin{equation}
\label{ex11} D(v)=4\pi^{2}m f_{0}(v)\int_{0}^{+\infty} dk
{k\hat{u}(k)^{2}\over |\epsilon(k,kv)|^2}.
\end{equation}
We note that the distribution function $P(v,t)$ of the test particle
relaxes towards the distribution of the bath $f_0(v)$ on a timescale
of the order  $N t_D$, where $t_D$ is the dynamical time \cite{hb}.
If we neglect collective effects, or consider the limit of large
velocities, we find that the diffusion coefficient is given by
\begin{equation}
\label{ex11bis} D(v)=4\pi^{2}m f_{0}(v)\int_{0}^{+\infty} dk
{k\hat{u}(k)^{2}}.
\end{equation}
It is proportional to the distribution function of the bath
$f_{0}(v)$. In particular, if the field particles are at statistical
equilibrium with a Gaussian distribution, the diffusion coefficient
decreases like $e^{-\beta m {v^{2}/ 2}}$. This type of Fokker-Planck
equations apply for example to the HMF model \cite{bouchet,bd,cvb}
which can be viewed as the one Fourier component of a one-dimensional
plasma (or self-gravitating system) \cite{cvb}. More generally, these
Fokker-Planck equations (\ref{ex10}) are valid for a wide class of one
dimensional systems with long range interactions \cite{hb}. We note
that for one-dimensional systems the Lenard-Balescu collision term
cancels out so that the distribution function of the field particles
$f(v,t)$ does not evolve, i.e.  $\partial f/\partial t=0$, on a
timescale of order $N t_D$. Since, on the other hand, the relaxation
time of a test particle toward the distribution of the bath is of
order $N t_D$, this implies that we can assume that the distribution
of the field particles is stationary $f(v,t)=f_0(v)$ when we study the
relaxation of a test particle. This is true for {\it any} distribution
function $f_0(v)$ that is a stable stationary solution of the Vlasov
equation \cite{bd,cvb,hb}. This is not true in higher dimensions $d=2$
and $d=3$, except for the Maxwellian distribution $f_e({\bf v})$,
since the distribution of the field particles $f({\bf v},t)$ changes
on a time $N t_D$ as it relaxes towards the statistical equilibrium
state $f_e({\bf v})$. Furthermore, even if we assume that the
distribution of the bath $f_{0}({\bf v})$ is approximately stationary,
the distribution of the test particle $P({\bf v},t)$ will not relax
towards $f_{0}({\bf v})$ for large times except if $f_{0}({\bf v})$ is
Maxwellian \cite{hb}.

A Fokker-Planck equation with a space dependent diffusion coefficient
has been introduced by Chavanis in
\cite{drift,kin,houches} to describe the relaxation of a test vortex
in a ``sea'' of field vortices with vorticity profile
$\omega_{0}(r)$. For an axisymmetric distribution, the Fokker-Planck
equation for $P(r,t)$ can be written \cite{kin,hb}:
\begin{eqnarray}
{\partial P\over\partial t}={1\over r}{\partial\over\partial
r}\biggl\lbrack rD(r)\biggl ({\partial P\over\partial r}-
P{d \over d r} \ln \omega_{0}\biggr )\biggr\rbrack,
\label{ex12}
\end{eqnarray}
with a diffusion coefficient
\begin{eqnarray}
D(r)={\gamma\over 8}{1\over |\Sigma(r)|}\ln N \omega_{0}(r),
\label{ex13}
\end{eqnarray}
where $\Sigma(r)=r\Omega'_{0}(r)$ is the local shear created by the
field vortices ($\Omega_{0}(r)$ represents the angular velocity
related to the vorticity by
$\omega_{0}(r)=(1/r)(r^{2}\Omega_{0})'$). For a vorticity profile
$\omega_{0}(r)=Ae^{-\lambda r^{2}}$ of the field vortices, it is easy
to see that the diffusion coefficient of the test vortex decreases
like $D(r)\sim r^{2}e^{-\lambda r^{2}}$ for $r\rightarrow +\infty$
\cite{hb,kinlemou}.

Finally, Fokker-Planck equations with diffusion and friction coefficients depending on the velocity can occur in many areas of physics. For example, the motion of atoms in a one-dimensional optical lattice formed by two counter propagating laser beams with linear perpendicular polarization can be described, after spatial averaging, by a Fokker-Planck of the form \cite{mark,lutz}:
\begin{eqnarray}
{\partial W\over\partial t}={\partial\over\partial p}\biggl\lbrack D(p){\partial W\over\partial p}-W K(p)\biggr\rbrack,
\label{ex14}
\end{eqnarray}
with 
\begin{eqnarray}
K(p)=-{\alpha p\over 1+(p/p_{c})^{2}}, \qquad D(p)=D_{0}+{D_{1}\over 1+(p/p_{c})^{2}}.
\label{ex15}
\end{eqnarray}
For $p\rightarrow +\infty$, $D(p)\rightarrow D_{0}$ and $K(p)\sim
-\alpha p_{c}^{2}/p$. This corresponds to a logarithmic potential
$U(p)\sim (\alpha p_{c}^{2}/D_{0})\ln p$ defined by
$K(p)/D(p)=-U'(p)$. Equation (\ref{ex14}) belongs to the general class
of Fokker-Planck equations that we shall study in the sequel. The case
of a logarithmic potential is treated specifically in
Sec. \ref{sec_log}.

\section{General solution of the problem}
\label{sec_gen}

The various examples discussed previously prompt us to study
Fokker-Planck equations with a diffusion coefficient and friction
force depending on the velocity. In particular, we can wonder how the
distribution function $f(v,t)$ approaches the equilibrium
distribution. This problem has been investigated by Potapenko {\it et
al.} \cite{potapenko} in the case of 3D plasmas where the diffusion
coefficient is a power law. These authors found that the asymptotic
behavior of the velocity distribution tail has a propagating wave
appearance. The high velocity tail develops a front at which the
distribution function drops to zero.  This front $v_{f}(t)$ progresses
with time and goes to $v_{f}(t)\rightarrow +\infty$ for $t\rightarrow
+\infty$. The profile of the front also deforms itself as time goes
on. However, if we use an appropriate system of coordinates, it can be
expressed in terms of the error function. We shall here formulate the
problem in a general setting, for an arbitrary form of diffusion
coefficient and friction force, and we shall investigate how the
evolution of the front and the profile of the high velocity tail
distribution depend on the form of the diffusion coefficient.

Let us consider the Fokker-Planck equation
\begin{eqnarray}
{\partial f\over\partial t}={1\over v^{d-1}}{\partial\over\partial
v}\biggl\lbrack v^{d-1}D(v)\biggl ({\partial f\over\partial v}+f
U'(v)\biggr )\biggr\rbrack,
\label{gen1}
\end{eqnarray}
where $D(v)\ge 0$ and $U(v)$ are arbitrary 
functions. If  the following zero flux  condition
\begin{equation}
\label{zeroflux}
v^{d-1}D(v)\left({\partial f\over\partial v}+f
U'(v)\right)  \rightarrow 0
\end{equation}
when $v\rightarrow \infty$ is fulfilled, then the stationary solutions of this Fokker-Planck
equation take the form
\begin{eqnarray}
f_{e}(v)=A e^{-U(v)},
\label{gen2}
\end{eqnarray}
where $A$ is a constant (normalization). The Fokker-Planck equation
(\ref{gen1}) decreases the free energy $F=E-S$ where $E=\int
f U(v)d{\bf v}$ is the energy and $S=-\int f\ln f d{\bf v}$ is the
Boltzmann entropy (the temperature has been included in the potential
$U$). Indeed, one has
\begin{eqnarray}
\dot F=-\int {D\over f}\biggl ({\partial f\over\partial {\bf v}}+f{\partial U\over\partial {\bf v}}\biggr )^{2}d{\bf v}\le 0.
\label{dotF}
\end{eqnarray}
Therefore, if $F$ is bounded from below, the distribution will converge towards the equilibrium state (\ref{gen2}) for $t\rightarrow +\infty$.
We want to analyze the
propagation of the front in the high velocity tail of the distribution
function. Thus, we set
\begin{eqnarray}
f(v,t)=f_{e}(v)u(v,t).
\label{gen3}
\end{eqnarray}
For sufficiently large times, the core of the distribution function
will have reached its asymptotic value (\ref{gen2}) so that $u\simeq 1$ in that
region. On the other hand, for sufficiently large velocities, the
distribution has not relaxed yet and $u=0$. Therefore, we expect the
formation of a front at a typical velocity value $\sim v_{f}(t)$ where
the function $u(v,t)$ passes from $u=1$ to $u=0$. On this
phenomenological basis, $u(v,t)$ is the relevant function to consider
in our ``travelling front'' analysis. Its exact evolution is governed by
\begin{eqnarray}
{\partial u\over\partial t}={1\over v^{d-1}}{\partial\over\partial v}\biggl ( v^{d-1}D(v){\partial u\over\partial v}\biggr )-D(v)U'(v){\partial u\over\partial v},
\label{gen4}
\end{eqnarray}
which is obtained from Eqs. (\ref{gen1})-(\ref{gen3}). If we perform
the change of variables $dx/dv=1/\sqrt{D(v)}$, we obtain the
equivalent equation
\begin{eqnarray}
{\partial u\over\partial t}={\partial^{2}u\over\partial x^{2}}+\biggl\lbrack {1\over 2}{D'(v)\over \sqrt{D(v)}}+{d-1\over v}\sqrt{D(v)}-U'(v)\sqrt{D(v)} \biggr\rbrack {\partial u\over\partial x},
\label{gen5}
\end{eqnarray}
where $v=v(x)$ must be viewed as an implicit function of $x$.
If we introduce the
velocity field
\begin{eqnarray}
V(v)=\sqrt{D(v)}\biggl \lbrack U'(v)-{d-1\over v}-{1\over 2}(\ln D)'(v)\biggr\rbrack,
\label{gen6}
\end{eqnarray}
or, equivalently,
\begin{eqnarray}
V(v)=-\sqrt{D(v)}{d\over d v}\biggl\lbrace \ln\biggl \lbrack v^{d-1}e^{-U(v)}D^{1/2}(v)\biggr\rbrack\biggr\rbrace,
\label{gen7}
\end{eqnarray}
Eq. (\ref{gen5}) can be re-written as
\begin{eqnarray}
{\partial u\over\partial t}+V(v) {\partial u\over\partial x}={\partial^{2}u\over\partial x^{2}}.
\label{gen8}
\end{eqnarray}
The structure of this equation is clear. The right hand side
corresponds to a diffusion and the left hand side
corresponds to an advection by a velocity field $V(v)$. This velocity
field (in velocity space) governs the evolution of the front. To
get a physical insight into the problem, let us first  neglect the
diffusion term. The resulting equation
\begin{eqnarray}
{\partial u\over\partial t}+V(v) {\partial u\over\partial x}=0
\label{gen9}
\end{eqnarray}
can be solved with the method of characteristics. Writing
\begin{eqnarray}
{dx\over dt}=V(v)={dx\over dv}{dv\over dt}={1\over\sqrt{D(v)}}{dv\over dt},
\label{gen10}
\end{eqnarray}
we get
\begin{eqnarray}
{dv\over dt}=\sqrt{D(v)}V(v)=D(v)\biggl \lbrack U'(v)-{d-1\over v}-{1\over 2}(\ln D)'(v)\biggr\rbrack.
\label{gen11}
\end{eqnarray}
This equation determines the evolution of the front $v_{f}(t)$.  In
the extreme approximation where the diffusion is neglected, the
profile of the front is given by a step function
$u(x,t)=\eta(x-x_{f}(t))$ where $\eta$ is the Heaviside function ($\eta(x)=1$ for $x<0$ and $\eta(x)=0$ for $x>0$).
 As we shall see, the diffusion term will smooth out this profile.

To take into account the effect of diffusion, we return to Eq. (\ref{gen8}) and
perform the change of variables
\begin{eqnarray}
z={x-x_{f}(t)}, \qquad u(x,t)=\phi(z,t),
\label{gen12}
\end{eqnarray}
where the function $x_{f}(t)$ is {\it defined} by Eq. (\ref{gen10}).
Substituting this in Eq. (\ref{gen8}), we obtain
\begin{eqnarray}
{\partial\phi\over\partial t}={\partial^{2}\phi\over\partial z^{2}}-\lbrack V(v)-V(v_{f})\rbrack {\partial\phi\over\partial z}.
\label{gen13}
\end{eqnarray}
So far, no approximation has been made so that Eq. (\ref{gen13}) is
exact and bears the same information as the initial Fokker-Planck
equation (\ref{gen1}). It is just written in a more convenient form
which will allow us to examine the situation in the region of the
front. Indeed, far from the front the profile is stationary (for
sufficiently long time) and Eq. (\ref{gen13}) is automatically
satisfied as $\phi=1$ for $v\ll v_f(t)$ and $\phi=0$ for $v\gg
v_f(t)$. If we consider values of the velocity that are close to
$v_{f}(t)$, we can expand the term in brackets in Taylor series (the
validity of this approximation will be studied in Sec.
\ref{sec_val}). Keeping only the first term in this expansion, we
get
\begin{eqnarray}
{\partial\phi\over\partial t}={\partial^{2}\phi\over\partial z^{2}}-g(t)z {\partial\phi\over\partial z},
\label{gen14}
\end{eqnarray}
where
\begin{eqnarray}
g(t)=V'(v_{f}(t))\sqrt{D(v_{f}(t))}.
\label{gen15}
\end{eqnarray}
For future convenience, we set  $\tau=2 t$ and define $h(\tau)=g(\tau/2)$. Therefore, the forgoing equation becomes
\begin{eqnarray}
{\partial\phi\over\partial \tau}={1\over 2}\biggl ({\partial^{2}\phi\over\partial z^{2}}-h(\tau)z {\partial\phi\over\partial z}\biggr ).
\label{gen16}
\end{eqnarray}
The general solution of this equation, for an arbitrary initial
condition, is given in Appendix \ref{sec_gensol}. Here, we look for
particular solutions of the form $\phi(z,\tau)=\Phi(z/\chi(\tau))$.
In Appendix \ref{sec_gensol}, we derive the condition under which
such solutions describe the asymptotic long time behavior of the
system (independently of the initial condition) and we check that
this condition is fulfilled for all the explicit examples that we
shall investigate in the following. To describe more general
situations, one must use the results of Appendix \ref{sec_gensol}.
Inserting the ansatz $\phi(z,\tau)=\Phi(z/\chi(\tau))$ in Eq.
(\ref{gen16}) leads to
\begin{eqnarray}
\Phi''+(2\chi\dot \chi-h \chi^{2})x\Phi'=0. \label{gen17}
\end{eqnarray}
The variables of velocity and time separate provided that the term
in parenthesis is a constant that we can arbitrarily set equal to
\begin{eqnarray}
2\chi\dot\chi-h \chi^{2}=2. \label{gen18}
\end{eqnarray}
Then,  $\Phi$ is the function
\begin{eqnarray}
\Phi(x)={1\over\sqrt{\pi}}\int_{x}^{+\infty}e^{-y^{2}}dy,
\label{gen19}
\end{eqnarray}
connected to the error function by $\Phi(x)={1\over 2}(1-{\rm
erf}(x))$. It satisfies $\Phi(-\infty)=1$, $\Phi(0)=1/2$ and
$\Phi(+\infty)=0$ so it reproduces the expected properties of the
front (it can be seen as a smooth step function). On the other hand,
solving for $\chi^{2}$ in Eq. (\ref{gen18}) and taking $\chi(1)=0$,
we finally obtain
\begin{eqnarray}
\chi^{2}(\tau)=2\int_{1}^{\tau}e^{\lbrack H(\tau)-H(\tau
  ')\rbrack}d\tau',
\label{gen20}
\end{eqnarray}
where $H$ is a primitive of $h$ with $H(1)=0$. The function
$\phi(z,\tau)=\Phi(z/\chi(\tau))$ with (\ref{gen19}) and
(\ref{gen20}) is the solution of Eq. (\ref{gen16}) for all times,
corresponding to the initial condition $\phi(z,1)=\eta(z)$ where
$\eta$ is a step function. It also governs the long time behavior of
the system for a large class of initial conditions (Appendix
\ref{sec_gensol}).

In particular, for $h(\tau)=\gamma/\tau$, we have
\begin{eqnarray}
{\partial\phi\over\partial \tau}={1\over 2}\biggl ({\partial^{2}\phi\over\partial z^{2}}-{\gamma\over\tau} z {\partial\phi\over\partial z}\biggr ),
\label{gen21}
\end{eqnarray}
and we recover the solution given in \cite{potapenko}, namely
\begin{eqnarray}
\phi(z,\tau)=\Phi\biggl\lbrack {z\over\sqrt{2}}\biggl ({1-\gamma\over \tau-\tau^{\gamma}}\biggr )^{1/2}\biggr\rbrack .
\label{gen22}
\end{eqnarray}
For $\gamma=-1$, a value that will frequently occur in the following
examples, we have
\begin{eqnarray}
\phi(z,\tau)=\Phi\biggl ( {z\over\tau^{1/2}}\biggr ), \qquad {\rm
for}\ \tau\gg 1. \label{gen23}
\end{eqnarray}
In terms of the function $g$, these analytical solutions correspond to
$g(t)=\lambda/t$. We come back to the original variables by setting
$\tau=2 t$ and $\gamma=2\lambda$.

\section{Condition of validity}
\label{sec_val}

We shall now investigate in greater detail the ability of our
approach to describe the front structure. First, we note that,
within our approximations,
\begin{eqnarray}
u(v_{f}(t),t)=\Phi(0)={1\over 2}, \label{val1}
\end{eqnarray}
so that $v_{f}(t)$ gives the position of the half-height profile. On the other hand, noting that
\begin{eqnarray}
V(v)-V(v_{f})=V'(v_{f})(v-v_{f})+{1\over 2}V''(v_{f})(v-v_{f})^{2}+...
\label{val2}
\end{eqnarray}
for $v\rightarrow v_{f}$, our approximation (\ref{gen14}) will be valid in
the range of velocities where we can neglect the
second term in the Taylor expansion with respect of the first. This
corresponds to
\begin{eqnarray}
|v-v_{f}|\ll \left|{2V'(v_{f})\over V''(v_{f})}\right|.
\label{val3}
\end{eqnarray}
In this range of velocities, our approach is accurate. Now, it will
provide a good description of the whole front if this range
$|v-v_{f}|$ is larger than the typical front half-width
$\Delta_{f}(t)/2$. This condition can be written
\begin{eqnarray}
\Delta_{f}(t)\ll \left|{4V'(v_{f})\over V''(v_{f})}\right|.
\label{val4}
\end{eqnarray}
Now, the front width can be estimated by
\begin{eqnarray}
\Delta_{f}(t)=\left|{\partial u\over\partial v}(v_{f}(t),t)\right|^{-1}.
\label{val5}
\end{eqnarray}
Within our approximation, this can be finally rewritten as
\begin{eqnarray}
\Delta_{f}(t)=\sqrt{\pi} \chi (2t) \sqrt{D(v_{f}(t))}. \label{val6}
\end{eqnarray}
Therefore, our approach will provide a good description of the front structure if
\begin{eqnarray}
\epsilon (t)\equiv \left| {\sqrt{\pi} \over
4}{\chi(2t)\sqrt{D(v_{f})}\over V'(v_{f})/V''(v_{f})}\right|\ll 1.
\label{val7}
\end{eqnarray}
Since we shall be interested by the large time limit, it is
particularly important to know the asymptotic behavior of the function
$\epsilon (t)$ for $t\rightarrow +\infty$. This has to be considered
case by case (see Sec. \ref{sec_part}).

The other assumption made in our study is that the long time
behavior of the system is described by
$\phi(z,\tau)=\Phi(z/\chi(\tau))$ with (\ref{gen19}) and
(\ref{gen20}). In Appendix \ref{sec_gensol}, we show that this is
the case if $\phi(z,1)\rightarrow 1$ for $z\rightarrow -\infty$, $\phi(z,1)\rightarrow 0$ for $z\rightarrow +\infty$ and if
\begin{eqnarray}
H_2(\tau)=\int_{1}^{\tau}e^{-H(\tau')}d\tau'\rightarrow +\infty \label{critG}
\end{eqnarray}
for $\tau\rightarrow +\infty$. In particular, for $h(\tau)=\gamma/\tau$, we have
\begin{eqnarray}
H(\tau)=\gamma\ln\tau, \qquad H_{1}(\tau)={1\over \tau^{\gamma/2}}, \label{brown1bis}
\end{eqnarray}
and 
\begin{eqnarray}
\chi^2(\tau)={2\over
1-\gamma}(\tau-\tau^{\gamma}), \qquad H_2(\tau)={\tau^{1-\gamma}-1\over 1-\gamma}, \label{brown1bisbis}
\end{eqnarray}
so that the criterion (\ref{critG}) is met if $\gamma\le 1$ and not
met if $\gamma>1$.

\section{Particular examples}
\label{sec_part}

We shall now discuss explicitly some particular examples of physical
interest and compare our theoretical predictions with direct numerical
simulations of the Fokker-Planck equation.

\subsection{Quadratic potential }
\label{sec_quad}

We first consider the case of a quadratic potential $U(v)=v^{2}/2$ leading to the Kramers equation
\begin{eqnarray}
{\partial f\over\partial t}={1\over v^{d-1}}{\partial\over\partial
v}\biggl\lbrack v^{d-1}D(v)\biggl ({\partial f\over\partial v}+f v\biggr )\biggr\rbrack,
\label{quad1}
\end{eqnarray}
with a variable diffusion coefficient $D(v)$. If we assume a zero flux
condition (\ref{zeroflux}), then the stationary solutions
are the  Maxwellian distributions
\begin{eqnarray}
f_{e}(v)=A e^{-{v^{2}\over 2}}.
\label{quad2}
\end{eqnarray}
We now consider various forms of diffusion coefficient and use the
general theory developed in Sec. \ref{sec_gen} to characterize the
front profile and its evolution. We shall only give asymptotic
expressions which are valid for large velocities and large time. This
will be implicit in all the following calculations.

\subsubsection{Power-law decay}
\label{sec_pw}

For the diffusion coefficient
\begin{eqnarray}
D(v)\sim v^{-\alpha}, \ \ \ \ \alpha >0, \label{pw1}
\end{eqnarray}
we get
\begin{eqnarray}
x\sim {2\over \alpha+2}v^{\alpha+2\over 2},\qquad v_{f}(t)\sim
(\alpha t)^{1/\alpha},  \label{pw2}
\end{eqnarray}
and
\begin{equation}
\left\{ \begin{array}{l} \ds g(t)\sim {2-\alpha\over 2\alpha}{1\over
t}, \ \ \mbox{if} \ \
\alpha\ne 2 \\
\ds g(t) \sim  {d-2 \over 2} {1\over t^2}, \ \ \mbox{if} \ \ \alpha
= 2.
\end{array}\right.
\end{equation}
For $\alpha=2$, we have assumed that the subdominant corrections to
the $D(v)\sim v^{-2}$ behavior are of order $v^{-4}$ or smaller.
Note that if $\alpha =2$ and $d=2$, then $g=0$ and the Fokker-Planck 
equation can be solved exactly (see Sec. \ref{sec_exact}).

For $\alpha\neq 2$, we find that $h=\gamma/\tau$ with
$\gamma=2/\alpha-1$. The front profile is given by
\begin{eqnarray}
u(v,t)\sim \Phi\biggl\lbrack {v^{\alpha+2\over 2}-(\alpha t)^{\alpha
+2 \over 2 \alpha}\over (\alpha+2) t^{1/2}}\biggl ({1-\gamma\over
1-(2t)^{\gamma-1}}\biggr )^{1/2}\biggr\rbrack. \label{pw3}
\end{eqnarray}
The criterion (\ref{critG}) is fulfilled only for $\alpha\ge 1$, so
that the above function provides the correct asymptotic behavior of
the solution, for any initial condition, only in that case. For
$\alpha<1$, it can however provide the correct asymptotic behavior if
the initial condition $\phi(z,1)$ is a step function (see Appendix
\ref{sec_gensol}).  Equation (\ref{pw3}) returns the results obtained
in \cite{potapenko}.  Of course this formula is written for $\alpha
\ne 1$, but it also provides the expression of the solution for
$\alpha=1$ by passing to the limit $\alpha \rightarrow 1$ yielding:
\begin{eqnarray}
u(v,t)\sim \Phi\left[ {v^{3\over 2}-t^{3
    \over 2 }\over 3(t\ln (2t) )^{1/2}}\right].
\label{p1}
\end{eqnarray}
On the other hand, for  $\alpha=2$ we find that $h=2(d-2)/\tau^2$
yielding $\chi^2(\tau)\sim 2\tau$ and $H_{2}(\tau)\sim \tau$ for
$\tau\rightarrow +\infty$. We find that the criterion (\ref{critG})
is satisfied and that the front profile is given by
\begin{eqnarray}
u(v,t)\sim \Phi\biggl\lbrack {v^{2}-2 t\over 4t^{1/2}}\biggr\rbrack,
\label{pw3br}
\end{eqnarray}
which turns out to be consistent with Eq. (\ref{pw3}).

\begin{figure}
\centerline{ \psfig{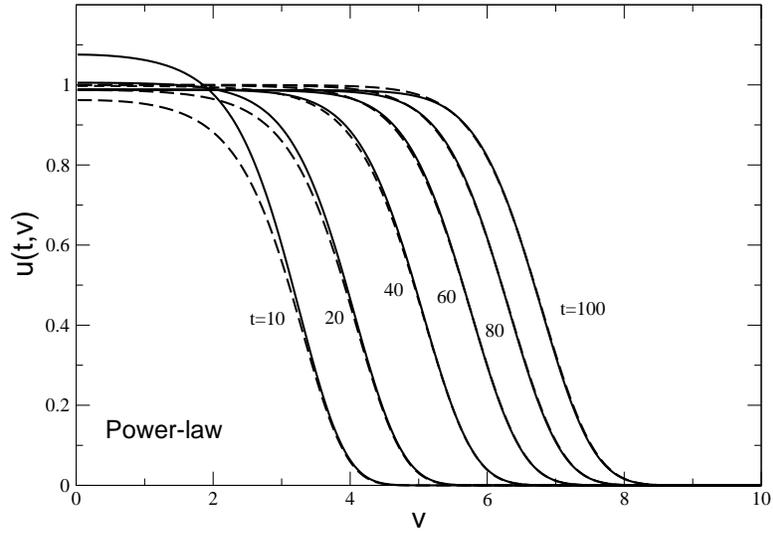}}
\caption{\label{profpuis} Evolution of the front profile $u(v,t)$
for a power-law diffusion coefficient with $\alpha=3$. For
sufficiently large times, we get a perfect agreement with the
theoretical profile (\ref{pw3}).  Here and in the following, the
straight lines correspond to the numerical simulation and the dashed
lines to the theoretical prediction. }
\end{figure}

\begin{figure}
\vskip1cm \centerline{
\psfig{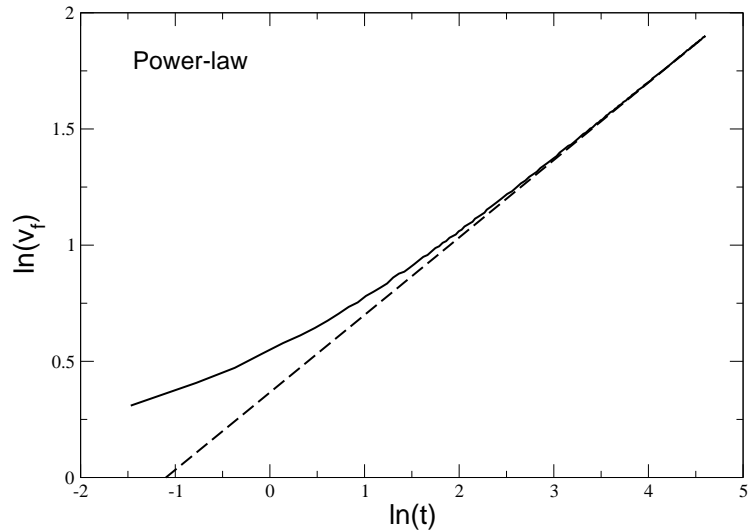}}
\caption{\label{frontpuis} Evolution of the front position
$v_{f}(t)$ defined by Eq. (\ref{val1}) for a power-law diffusion
coefficient with $\alpha=3$. For sufficiently large times, it
coincides with the theoretical prediction (\ref{pw2})-b.}
\end{figure}

Concerning the validity of this approach in relation with the
criterion (\ref{val7}), we have
\begin{equation}
\left\{\begin{array}{l}
\ds
 \epsilon (t)\sim  A_\alpha  t^{-1/\alpha},  \ \ \ \mbox{if}\ \
 \alpha>1 \ \ \mbox{and} \ \ {\alpha \ne 2}, \\
\ds
 \epsilon (t)\sim  {B_\alpha \over  t}  \ \ \ \mbox{if}\ \
\alpha<1,\\ 
\ds
 \epsilon (t)\sim {\sqrt{\pi} \over
 4}{(\ln t)^{1/2}\over t} \ \ \ \mbox{if} \ \ \alpha=1, \\
\ds
 \epsilon (t)\sim {3\sqrt{\pi} \over
 4}t^{-1/2} \ \ \ \mbox{if} \ \ \alpha=2 \ \ (d\neq 2),
\end{array}\right.
\end{equation}
where $A_\alpha= (\sqrt{2\pi}/8)\alpha ^{1-1/\alpha} (\alpha
-1)^{-1/2}$ and $ B_\alpha= (\sqrt{\pi}/8)\alpha ^{1-1/\alpha}
(1-\alpha)^{-1/2} 2^{1/\alpha -1/2}$.  Therefore, the condition
$\epsilon\ll 1$ is always fulfilled for sufficiently large times. For
sake of illustration, we show the case $\alpha=3$ for $d=3$ (plasmas
and stellar systems) in Figs. \ref{profpuis} and
\ref{frontpuis}. These results are obtained by solving numerically the Fokker-Planck equation (\ref{quad1}) starting from a step
function: $f(v,t=0)=3/4\pi$ if $v<1$ and $f(v,t=0)=0$ if $v>1$ (water-bag). In
the simulation we have adopted the expression (\ref{ex6})-a of the diffusion
coefficient which reduces to Eq. (\ref{pw1}) with $\alpha=3$ for $v\rightarrow
+\infty$. Other examples are given in \cite{potapenko}.

\subsubsection{Gaussian decay}
\label{sec_g}

We now consider a diffusion coefficient of the form
\begin{eqnarray}
D(v)\sim e^{-{\gamma v^{2}}}. \label{g1}
\end{eqnarray}
Considering  the evolution of the high velocity tail, we have 
\begin{eqnarray}
x\sim \int_{0}^{v} e^{{\gamma\over 2}y^{2}}dy. \label{g2}
\end{eqnarray}
For large $v$, we obtain the relation (see Appendix \ref{sec_ab})
\begin{eqnarray}
x\sim {1\over \gamma v}e^{{\gamma\over 2}v^{2}}.
\label{g3}
\end{eqnarray}
The  position of the front $v_{f}(t)$ is determined by
\begin{eqnarray}
\int^{\gamma v_{f}^{2}}{e^{y}\over y}dy\sim 2(\gamma+1)t.
\label{g4}
\end{eqnarray}
The integral can be expressed in terms of the exponential integral
$E_{i}(x)$. For large times, we get
\begin{eqnarray}
{e^{\gamma v_{f}^{2}}\over \gamma v_{f}^{2}}\sim 2(\gamma+1)t.
\label{g5}
\end{eqnarray}
To leading order, we have
\begin{eqnarray}
v_{f}(t)\sim \biggl ({\ln t\over\gamma}\biggr )^{1/2}.
\label{g6}
\end{eqnarray}
We note that the evolution of the front is extremely slow
(logarithmic). On the other hand, we find that
\begin{eqnarray}
g(t)\sim -{1\over 2t}, \label{g7}
\end{eqnarray}
implying that the criterion  (\ref{critG}) is fulfilled. In order to
determine the front profile, we need first to evaluate $x_{f}(t)$.
An equivalent for $t\rightarrow +\infty$ is obtained by combining
Eqs. (\ref{g3}) and (\ref{g5}) leading to
\begin{eqnarray}
x_{f}(t)\sim \sqrt{2(\gamma+1)t\over \gamma}.
\label{g8}
\end{eqnarray}
Therefore, the front profile is given by
\begin{eqnarray}
u(v,t)\sim \Phi\biggl\lbrack {{1\over \gamma v}e^{{\gamma\over
2}v^{2}}-\sqrt{{2(\gamma+1)t\over \gamma}}\over
\sqrt{2t}}\biggr\rbrack. \label{g9}
\end{eqnarray}
Concerning the validity of this approach in relation with the
criterion (\ref{val7}), we find that $\epsilon(t)\rightarrow {1\over
4}({\pi\gamma\over\gamma+1})^{1/2}$ for $t\rightarrow +\infty$ so
that our approximations are marginally valid: $\epsilon(t)$ does not
go to zero but it does not diverge with time neither.

\begin{figure}
\centerline{
\psfig{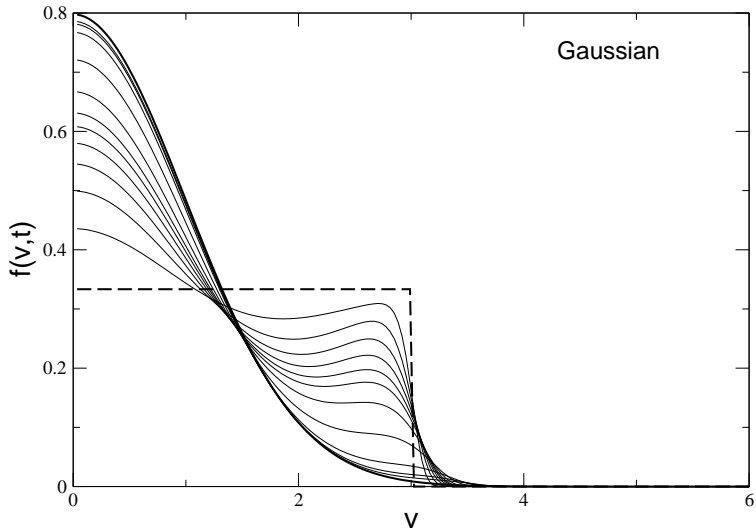}}
\caption{\label{shortt} Short time evolution of the distribution
function $f(v,t)$ for a Gaussian diffusion coefficient. This figure
shows the slow depletion of the high velocity tail due to the rapid
decrease of the diffusion coefficient for $v\gg v_{th}=1$. }
\end{figure}

We have performed numerical simulations of the Fokker-Planck
equation (\ref{quad1}) with the diffusion coefficient (\ref{g1})
with $\gamma=1/2$.  As discussed in Sec. \ref{sec_ex}, this equation
describes the evolution of a ``test particle'' in a bath of ``field
particles'' at statistical equilibrium (with Maxwellian distribution
function $f_{0}\sim e^{-v^{2}/2}$) for one-dimensional systems
($d=1$) with long-range interactions such as the HMF model. In Fig.
\ref{shortt}, we show the ``short'' time evolution of the
distribution function $f(v,t)$ starting from a step function:
$f(v,0)=1/3$ for $v\le 3$ whose width is larger than the thermal
speed $v_{th}=1$ (dashed line).  Since $D(v)$  rapidly decreases with
the velocity for $v\gg v_{th}=1$, the high velocity component of the
initial condition takes time to be depleted. Indeed, the core of the
distribution rapidly reaches a Maxwellian distribution while the
tail keeps the memory of the initial condition and remains flat for
relatively long times. For the parameters of the simulation, this
can even create a non-monotonic distribution for short times as
shown in Fig. \ref{shortt} (see Appendix \ref{sec_mono}).

\begin{figure}
\centerline{
\psfig{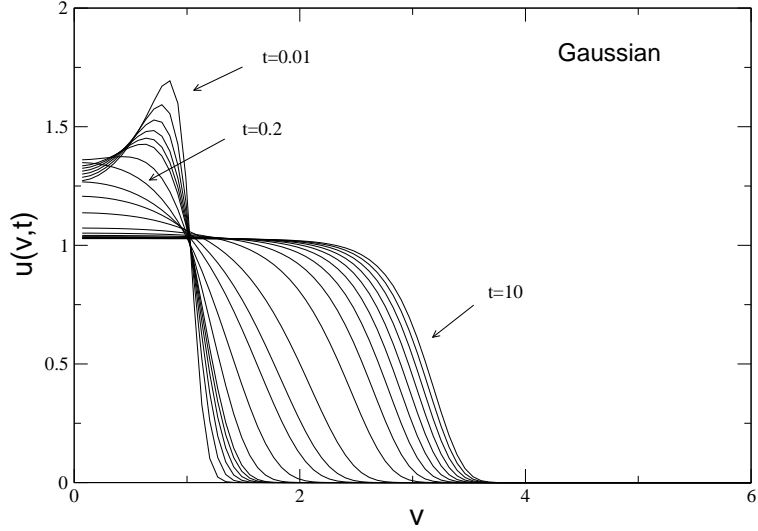}}
\caption{\label{smallu} Evolution of the normalized distribution function $u(v,t)$ for a Gaussian diffusion coefficient for short and large times. For short times the normalized distribution function forms a bump which slowly disappears as the core of the distribution becomes Maxwellian. For large times the function $u(v,t)$ has a front structure.  }
\end{figure}

\begin{figure}
\centerline{ \psfig{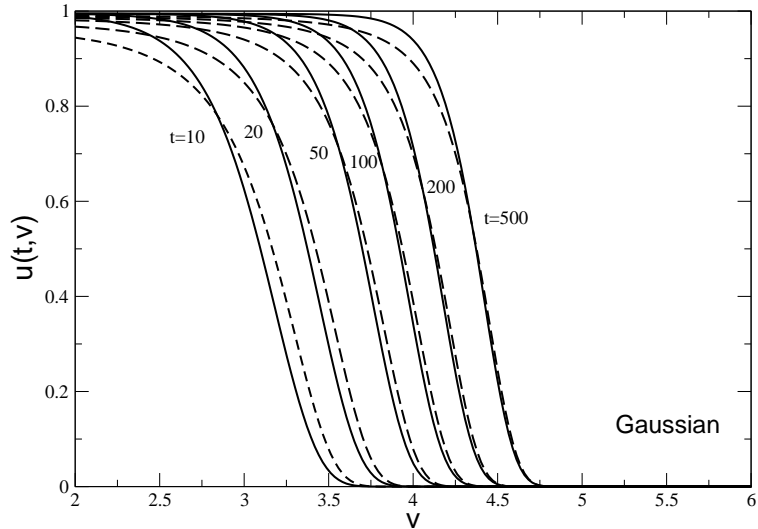}}
\caption{\label{profgauss} Evolution of the front profile $u(v,t)$
for a Gaussian diffusion coefficient. For sufficiently large times,
we get a fair agreement with the theoretical profile (\ref{g9})
although the validity of our approach was expected to be marginal in
that case. }
\end{figure}

\begin{figure}
\centerline{
\psfig{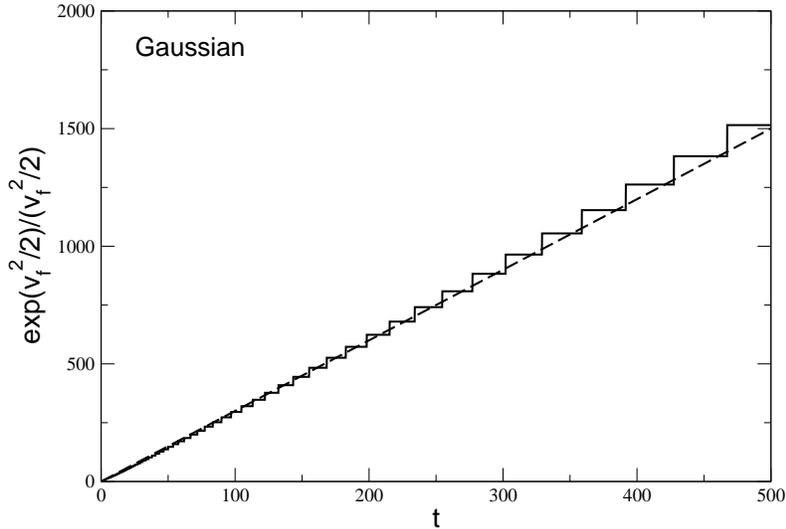}}
\caption{\label{frontgauss}Evolution of the front position $v_{f}(t)$ defined by Eq. (\ref{val1}) for a Gaussian diffusion coefficient. For sufficiently large times, we get a fair agreement with the theoretical prediction  (\ref{g5}).}
\end{figure}

In Fig. \ref{smallu} we show the evolution of the normalized
distribution function $u(v,t)$ for short and large times. In that
case, we start from an initial condition: $f(v,0)=1$ if $v<1$ and
$f(v,0)=0$ if $v>1$. In Fig. \ref{profgauss}, we focus on the
evolution of the front for large times and we compare the result of
the numerical simulation with the theoretical prediction. In Fig.
\ref{frontgauss}, we compare the evolution of the front displacement
$v_{f}(t)$ with the theoretical prediction. The numerical results
are in fair agreement with the theory although its domain of
validity was expected to be marginal in that case according to our
estimates. This fair agreement may be explained by the relatively
small value of $\epsilon(+\infty)= {1\over 4}(\pi/3)^{1/2}\simeq
0.255...$ for $\gamma=1/2$ even if this parameter does not strictly
tends to zero for $t\rightarrow +\infty$. We note that for
$t\rightarrow +\infty$ and $v\ll v_{f}(t)$, we have $u\sim
\Phi(-\sqrt{(\gamma+1)/\gamma})=\Phi(-\sqrt{\pi}/4\epsilon)$ which
has not converged to $u=1$ precisely because $\epsilon\neq 0$.
However, for $\epsilon(+\infty)\simeq 0.255...$, we find that
$u\simeq 0.993$ which is very close to one.

In relation with the HMF model, we would like to recall that the
present approach describes the relaxation of a given test particle
immersed in a bath of field particles at statistical equilibrium. The
relaxation is due to finite $N$ effects (correlations). This does not
describe the evolution of the distribution function of the system as a
whole. The clearest reason is that the Fokker-Planck equation
(\ref{ex1}) resulting from a thermal bath approximation does not
conserve energy contrary to the microcanonical evolution of a
Hamiltonian system. In the collisionless regime, the distribution
function $f(\theta,v,t)$ of the HMF model is governed by the Vlasov
equation coupled to the mean-field potential $\Phi(\theta,t)=-{k\over
2\pi}\int\cos(\theta-\theta')f(\theta',v',t)d\theta'dv'$ produced by
the particles. The Vlasov equation can experience a process of violent
relaxation and converge toward a Quasi Stationary State (QSS) on the
coarse-grained scale \cite{qss}. Some dynamical theories of violent
relaxation \cite{csr} based on a Maximum Entropy Production Principle
(MEPP) propose to model the evolution of the coarse-grained
distribution function by a generalized Fokker-Planck equation of the
form (for a water-bag initial condition):
\begin{eqnarray}
\label{cg}
{\partial \overline{f}\over\partial t}+v{\partial \overline{f}\over\partial \theta}-\nabla\Phi {\partial \overline{f}\over\partial v}={\partial\over\partial v}\biggl\lbrace D(\theta,v,t)\biggl \lbrack{\partial \overline{f}\over\partial v}+\beta(t)\overline{f}(\eta_{0}-\overline{f}) v\biggr \rbrack\biggr\rbrace,
\end{eqnarray}
where $\beta(t)$ evolves so as to conserve energy (see \cite{csr} for
more details). An important point is that the diffusion coefficient is
not constant but is related to the correlations of the fine-grained
fluctuations. It can depend on position, velocity and time and can be
very small in certain regions of phase-space and for large times. The
vanishing or smallness of the diffusion coefficient can slow down the
dynamics and lead to a confinement of the distribution function in
phase space which may be qualitatively similar to what is shown in
Fig. \ref{shortt}. However, the dynamical equation (\ref{cg}) and its
physical interpretation are different from Eq. (\ref{ex10}), so that
their similarity is, at most, an analogy.

\subsubsection{Exponential decay}
\label{sec_e}

For the diffusion coefficient
\begin{eqnarray}
D(v)\sim e^{-\gamma v}, \label{e1}
\end{eqnarray}
we get
\begin{eqnarray}
x\sim {2\over \gamma}e^{{\gamma\over 2}v}. \label{e2}
\end{eqnarray}
The evolution of the front is given by
\begin{eqnarray}
{e^{\gamma v_{f}}\over\gamma v_{f}}\sim t.
\label{e3}
\end{eqnarray}
To leading order, we have
\begin{eqnarray}
v_{f}(t)\sim {1\over\gamma}\ln t,
\label{e4}
\end{eqnarray}
which shows that the progression of the front is again very slow. We also
have
\begin{eqnarray}
g(t)\sim -{1\over 2t}, \label{e5}
\end{eqnarray}
implying that the criterion  (\ref{critG}) is fulfilled. In order to
determine the front profile, we need first to evaluate $x_{f}(t)$.
An equivalent for $t\rightarrow +\infty$ is obtained by combining
Eqs. (\ref{e2}) and (\ref{e3}) leading to
\begin{eqnarray}
x_{f}(t)\sim {2\over\gamma} (t\ln t)^{1/2}.
\label{e6}
\end{eqnarray}
However, since the evolution with time is slow, we shall work with the
more precise expression $x_{f}(t)\sim {2\over\gamma} (t\ln (t\ln
t))^{1/2}$ obtained by keeping the term of next order. With this
expression, we find that the front profile is given by
\begin{eqnarray}
u(v,t)\sim \Phi\biggl\lbrack \sqrt{2}{e^{{\gamma\over 2}v}-(t\ln
(t\ln t))^{1/2}\over \gamma t^{1/2}}\biggr\rbrack. \label{e7}
\end{eqnarray}
Concerning the validity of this approach in relation with the
criterion (\ref{val7}), we find that $\epsilon(t)\sim {\gamma\over
8}\sqrt{2\pi}(\ln t)^{-1/2}$ for $t\rightarrow +\infty$ so that our
approximations are valid for sufficiently large times. Note that the
decay is slow with time (logarithmic) but the prefactor
${\gamma\over 8}\sqrt{2\pi}= 0.313...$ is relatively small (for
$\gamma=1$) which can explain the very good agreement with the
numerics even for moderate timescales. Figures \ref{profexp} and
\ref{frontexp} are obtained by solving the Fokker-Planck equation
(\ref{quad1}) with the diffusion coefficient (\ref{e1}) with
$\gamma=1$, starting from an initial condition: $f(v,0)=1$ if $v<1$
and $f=0$ if $v>1$.

\begin{figure}
\centerline{ \psfig{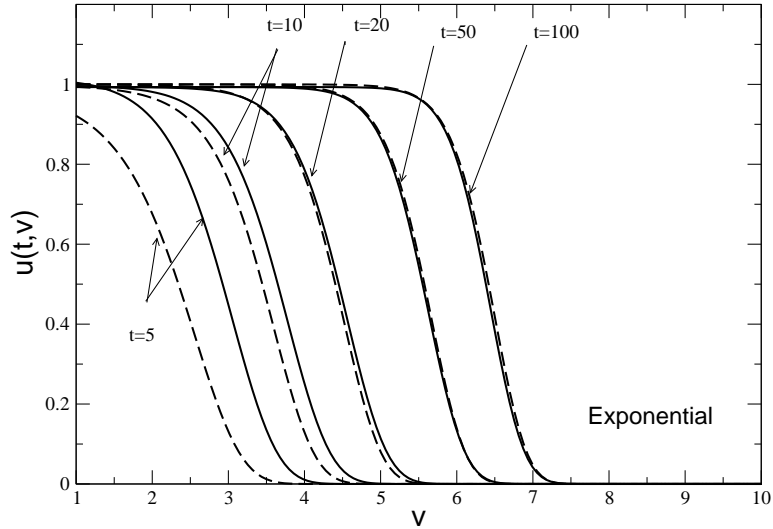}}
\caption{\label{profexp} Evolution of the front profile $u(v,t)$ for
an exponential diffusion coefficient. For sufficiently large times,
we get an excellent agreement with the theoretical profile
(\ref{e7}).}
\end{figure}

\begin{figure}
\centerline{
\psfig{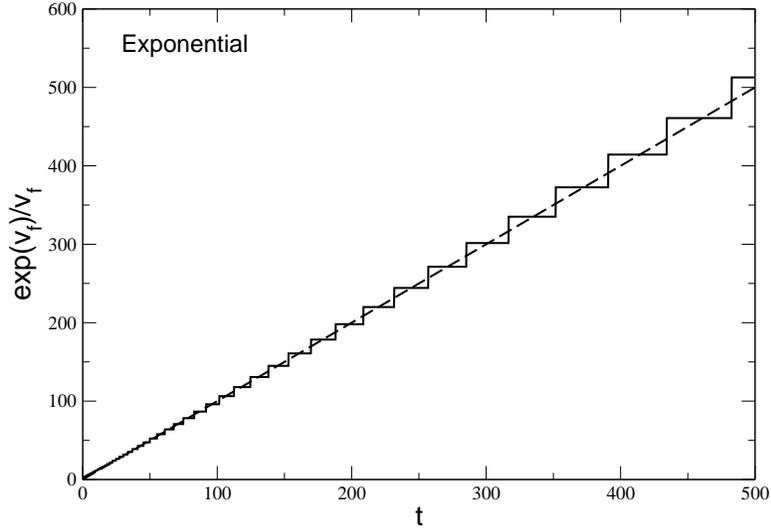}}
\caption{\label{frontexp} Evolution of the front position $v_{f}(t)$
defined by Eq. (\ref{val1}) for an exponential diffusion
coefficient. For sufficiently large times, we get an excellent
agreement with the theoretical prediction  (\ref{e3}). }
\end{figure}

\subsubsection{Stretched exponential}
\label{sec_se}

For the diffusion coefficient
\begin{eqnarray}
D(v)\sim e^{-\gamma v^{\delta}}, \ \ \ \delta >0, \label{se1}
\end{eqnarray}
we get
\begin{eqnarray}
x\sim \int_{0}^{v} e^{{\gamma\over 2}y^{\delta}}dy. \label{se2}
\end{eqnarray}
We shall briefly discuss how the results depend on the value of
$\delta$. A more detailed analysis can be carried out as in the
previous sections where the cases $\delta=1$ and $\delta=2$ are
explicitly considered.

For  $\delta<2$, the front evolution is given  by
\begin{eqnarray}
{e^{\gamma v_{f}^{\delta}}\over\gamma v_{f}^{\delta}}\sim \delta t.
\label{se3}
\end{eqnarray}
To leading order, we get
\begin{eqnarray}
v_{f}(t)\sim \biggl ({\ln t\over \gamma}\biggr )^{1/\delta}.
\label{se4}
\end{eqnarray}
On the other hand,
\begin{eqnarray}
g(t)\sim -{1\over 2 t}, \label{se5}
\end{eqnarray}
implying that the criterion  (\ref{critG}) is fulfilled. Concerning
the validity of this approach in relation with the criterion
(\ref{val7}), we find that
\begin{equation}
 \epsilon (t) \sim {\sqrt{2\pi \delta} \over 8}
\left(\ln t \over \gamma \right)^{-(2-\delta)/2\delta},
\end{equation}
 so that the criterion is satisfied for sufficiently
large times (but slowly).

For $\delta>2$, the front evolution is given by
\begin{eqnarray}
{e^{\gamma v_{f}^{\delta}}\over (\gamma v_{f}^{\delta})^{2(\delta-1)/\delta}}\sim {1\over 2}\gamma^{2/\delta}\delta^{2} t.
\label{se6}
\end{eqnarray}
To leading order, we get
\begin{eqnarray}
v_{f}(t)\sim \biggl ({\ln t\over \gamma}\biggr )^{1/\delta}.
\label{se7}
\end{eqnarray}
On the other hand,
\begin{eqnarray}
g(t)\sim -{1\over 2 t}, \label{se8}
\end{eqnarray}
implying that the criterion  (\ref{critG}) is fulfilled. Concerning
the validity of this approach in relation with the criterion
(\ref{val7}), we find that $\epsilon\rightarrow {\sqrt{\pi} \over
4}$ so that this approach is marginally valid.

\subsection{Linear potential}
\label{sec_lin}

We now consider the case of a linear
potential $U(v)=\gamma v$ leading to a Fokker-Planck equation of the
form
\begin{eqnarray}
{\partial f\over\partial t}={1\over v^{d-1}}{\partial\over\partial v}\biggl\lbrack v^{d-1}D(v)\biggl ({\partial f\over\partial v}+\gamma f \biggr )\biggr\rbrack.
\label{lin1}
\end{eqnarray}
The stationary solution is
\begin{eqnarray}
f_{e}(v)=Ae^{-\gamma v}.
\label{lin2}
\end{eqnarray}
For a diffusion coefficient decreasing algebraically with the velocity
\begin{eqnarray}
D(v)\sim v^{-\alpha}, \ \ \ \ \alpha >0,\label{lin3}
\end{eqnarray}
we get
\begin{eqnarray}
x\sim {2\over \alpha+2}v^{\alpha+2\over 2}, \qquad v_{f}(t)\sim
\lbrack (\alpha+1)\gamma t\rbrack^{1/(\alpha+1)},
\label{lin4}
\end{eqnarray}
and 
\begin{eqnarray}
 g(t)\sim
-{\alpha \over 2(\alpha+1)}{1\over t}, \label{lin4bis}
\end{eqnarray}
implying that the criterion  (\ref{critG}) is fulfilled.
We also find that $\epsilon(t)\propto t^{-1/(2(\alpha+1))}$ so that
the validity criterion of our approach is always fulfilled for
sufficiently large times.

\subsection{Logarithmic potential}
\label{sec_log}

In this subsection, we consider the case of a constant diffusion coefficient $D(v)=1$ and a logarithmic potential $U(v)=(\alpha/2)\ln(1+v^{2})$ in $d=1$ leading to a Fokker-Planck equation of the form
\begin{eqnarray}
{\partial f\over\partial t}={\partial\over\partial v}\biggl ({\partial f\over\partial v}+ \alpha f {v\over 1+v^{2}}\biggr ).
\label{log1}
\end{eqnarray}
This type of Fokker-Planck equations arises in the study of optical lattices \cite{mark,lutz}. The stationary solution of this equation is of the form
\begin{eqnarray}
f_{e}(v)={A\over (1+v^{2})^{\alpha/2}}.
\label{log2}
\end{eqnarray}
This is similar to a Tsallis distribution with $q=(\alpha-2)/\alpha$
\cite{lutz2}. However, it arises here from a linear Fokker-Planck
equation (associated with the Boltzmann entropy) with a logarithmic
potential, instead of a nonlinear Fokker-Planck equation (associated
with the Tsallis entropy) with a quadratic potential
\cite{pp,bukman,pre}. The distribution (\ref{log2}) is normalizable provided
that $\alpha>\alpha_{crit}=1$. The case $\alpha=2$ corresponds to the
Lorentzian. The evolution of the front $v_{f}(t)$ satisfies
\begin{eqnarray}
{dv_{f}\over dt}={\alpha v_{f}\over 1+v_{f}^{2}}.
\label{log3}
\end{eqnarray} 
The exact solution is given by $v_{f}^{2}e^{v_{f}^{2}}=e^{2\alpha t+C}$ where $C$ is a constant of integration. It can be written $v_{f}^{2}(t)=W(e^{2\alpha t+C})$ where $W(x)$ is the Lambert function which is solution of the transcendental equation $W(x){\rm exp}\lbrack W(x)\rbrack =x$.  For $t\rightarrow +\infty$, we get $v_{f}(t)\sim (2\alpha t)^{1/2}$ and $g(t)\sim -1/(2t)$. Therefore, the front profile is given by
\begin{eqnarray}
u(v,t)=\Phi\biggl ({v-\sqrt{2\alpha t}\over \sqrt{2t}}\biggr ),
\label{log4}
\end{eqnarray} 
for large times. Concerning the validity of our approach with respect
to the criterion (\ref{val7}), we find that $\epsilon\rightarrow
({\pi\over 4\alpha})^{1/2}$ for $t\rightarrow +\infty$ so that our
approach is marginally valid. The theoretical prediction is not very
good for $\alpha=2$ but the agreement improves for large values of
$\alpha\gg \alpha_{crit}=1$ for which $\epsilon$ is reduced. In
particular, the results of numerical simulations performed with
$\alpha=10$ are shown in Figs. \ref{frontLOG} and
\ref{profilsLOG}. Concerning the evolution of the front, we find a
good agreement with the Lambert function except that the measured
exponent in Fig. \ref{frontLOG} is $\sim 20.6$ instead of
$2\alpha=20$. Therefore, the front increases a little bit faster than
the theoretical prediction. This is confirmed in Fig. \ref{profilsLOG}
which shows the evolution of the front profile. We note, however, the
relatively good agreement between direct numerical simulation and
theory. The slight discrepency is due to the finite value of
$\epsilon\simeq 0.28$.

\begin{figure}
\centerline{ \psfig{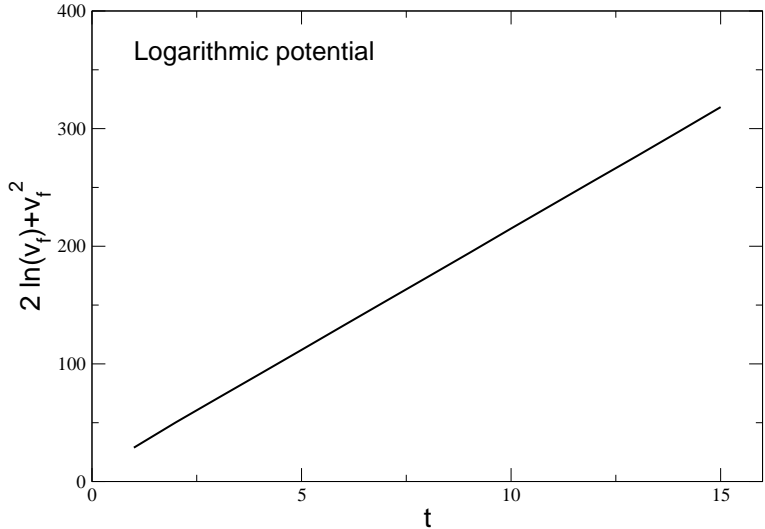}}
\caption{Evolution of the front position $v_{f}(t)$ for a logarithmic potential with $\alpha=10$. We get a fair agremment with the Lambert function.\label{frontLOG}}
\end{figure}

\begin{figure}
\centerline{ \psfig{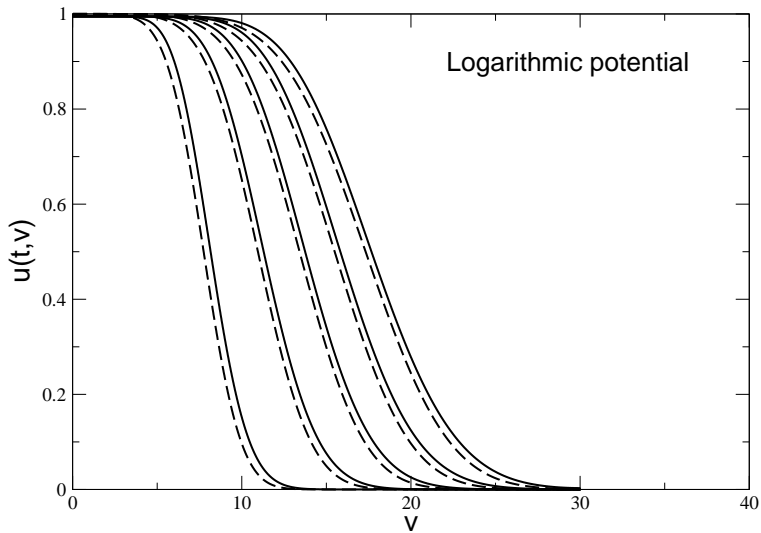}}
\caption{Evolution of the front profile $u(v,t)$  for a logarithmic potential with $\alpha=10$.\label{profilsLOG}}
\end{figure}

\subsection{Fokker-Planck equations associated with one-dimensional systems with long-range interactions}
\label{sec_one}

We now consider the case of Fokker-Planck equations describing one
dimensional systems with long-range interactions. In that case, the
diffusion coefficient $D\sim f_{0}(v)$ for $v\rightarrow +\infty$
and the potential $U(v)=-\ln f_{0}(v)$ are expressed in terms of the
distribution of the bath $f_{0}(v)$. The corresponding Fokker-Planck
equation can be rewritten as (see Sec. \ref{sec_ex}):
\begin{eqnarray}
{\partial f\over\partial t}={\partial\over\partial v}\biggl (f_{0}{\partial f\over\partial v}-f{df_{0}\over dv} \biggr ).
\label{one1}
\end{eqnarray}
The general relations obtained in Sec. \ref{sec_gen} take the simpler forms
\begin{eqnarray}
{dx\over dv}\sim {1\over\sqrt{f_{0}(v)}}, \qquad V(v)\sim -{3\over
2}{f_{0}'(v)\over \sqrt{f_{0}(v)}}, \qquad {dv_{f}\over dt}\sim
-{3\over 2}f_{0}'(v_{f}). \label{one2}
\end{eqnarray}
The case where $f_{0}(v)$ is a Gaussian distribution (thermal bath)
has already been studied in Sec. \ref{sec_g}. We shall consider
other examples where $f_{0}(v)$ is not necessarily the statistical
equilibrium state. This description still makes sense if
$f_{0}$ is a stable stationary solution of the Vlasov equation because
the relaxation of the system as a whole is slower than the relaxation
of a test particle in a fixed distribution.

\subsubsection{Exponential distribution}
\label{sec_oe}

For the exponential distribution
\begin{eqnarray}
f_{0}\sim e^{-\gamma v}, \label{oe1}
\end{eqnarray}
we obtain
\begin{eqnarray}
x\sim {2\over\gamma}e^{{\gamma\over 2}v},\qquad v_{f}\sim
{1\over\gamma}\ln \biggl ({3\over 2}\gamma^{2}t\biggr ), \label{oe2}
\end{eqnarray}
and
\begin{eqnarray}
g(t)\sim -{1\over 2t}, \label{oe2bis}
\end{eqnarray}
implying that the criterion  (\ref{critG}) is fulfilled. The front
profile is given by
\begin{eqnarray}
u(v,t)\sim \Phi\biggl\lbrack \sqrt{2}{e^{{\gamma\over 2}v}-({3\over
2}\gamma^{2}t)^{1/2}\over \gamma t^{1/2}}\biggr\rbrack.\label{oe3}
\end{eqnarray}
Concerning the validity of this approach in relation with the
criterion (\ref{val7}), we find that $\epsilon\rightarrow {1\over
4}({\pi\over 3})^{1/2}\simeq 0.256...$ for $t\rightarrow +\infty$ so
that our approximations are marginally valid and are accurate since
$\epsilon$ is relatively small.

\subsubsection{Power-law distribution}
\label{sec_opw}

For the power-law distribution
\begin{eqnarray}
f_{0}\sim v^{-\alpha}, \label{opw1}
\end{eqnarray}
we obtain
\begin{eqnarray}
x\sim {2\over\alpha+2}v^{\alpha+2\over 2},\qquad v_{f}(t)\sim
\biggl\lbrack {3\over
2}\alpha(\alpha+2)t\biggr\rbrack^{1\over\alpha+2},\label{opw2}
\end{eqnarray}
and
\begin{eqnarray}
 g(t)\sim
-{1\over 2t},\label{opw2bis}
\end{eqnarray}
implying that the criterion  (\ref{critG}) is fulfilled. The front
profile is given by
\begin{eqnarray}
u(v,t)\sim \Phi\biggl\lbrack \sqrt{2}{v^{\alpha+2\over
2}-\sqrt{{3\over 2}\alpha(\alpha+2)t}\over (\alpha+2)
t^{1/2}}\biggr\rbrack. \label{opw3}
\end{eqnarray}
Concerning the validity of this approach in relation with the
criterion (\ref{val7}), we find that $\epsilon\rightarrow {1\over
8}(\alpha+4)\sqrt{4\pi\over 3\alpha(\alpha+2)}$ for $t\rightarrow
+\infty$ so that our approximations are marginally valid.

\section{A class of exactly solvable Fokker-Planck equations}
\label{sec_exact}

In this section we  give a class of Fokker-Planck equations
(\ref{gen1}) for which our approach turns out to be exact. More
precisely, we derive a relationship between $D(v)$ and $U(v)$ under
which  the corresponding Fokker-Planck equation (\ref{gen1}) can be
exactly solved. First, we recall that equations (\ref{gen1}) and
(\ref{gen13}) are equivalent and no approximation has been made in
the derivation of  Eq. (\ref{gen13}) from Eq. (\ref{gen1}). However to pass
from Eq. (\ref{gen13}) to Eq. (\ref{gen14}) for  a general velocity field
$V(v)$, we have made an  approximation and have assumed  that the
concerned  velocities are close enough to the position of the front
$v_f(t)$.  This is in fact the only approximation in the approach
developed above. We now look for a situation where such an
approximation  is exact. This is the case if $V(v)=Ax+B$ where $A$
and $B$ are arbitrary constants. Using
\begin{equation}
{d \over d x} ( V(v)) = V'(v)\sqrt{D(v)} = A,
\end{equation}
and replacing $V$ by its expression (\ref{gen6}), we equivalently get
\begin{equation}
\sqrt{D(v)} {d \over d v}\left(\sqrt{D(v)}\left\lbrack
U'(v)-{d-1\over
  v}-{1\over 2}(\ln D)'(v)  \right\rbrack \right)=A.
\end{equation}
Defining $R(v)=\sqrt{D(v)}$, we obtain
\begin{equation}
R(v){d \over d v}\left[\left(U'(v)-{d-1\over
  v}\right) R -R'(v)\right] =A.
\label{relationDU1}
\end{equation}
If $U(v)$ and $R(v)=\sqrt{D(v)}$ satisfy Eq. (\ref{relationDU1}), as
discussed in Appendix \ref{sec_gdu}, we can use the results of
Appendix \ref{sec_gensol} with $g(t)=A$ to obtain exact solutions of
Eq. (\ref{gen1}) for all times.

Let us make these solutions more explicit. First of all, for
$h(\tau)=A$, we have
\begin{eqnarray}
H(\tau)=A(\tau-1), \qquad H_1(\tau)=e^{-{A\over 2}(\tau-1)},
\end{eqnarray}
\begin{eqnarray}
\chi^2(\tau)={2\over A}(e^{A(\tau-1)}-1), \qquad H_2(\tau)={1\over
A}(1-e^{-A(\tau-1)}). \label{exa1}
\end{eqnarray}
We note that the criterion (\ref{critG}) is met only for $A\le 0$. On the other hand, integrating
\begin{eqnarray}
{dx\over dt}=V(v)=Ax+B, \label{exa2}
\end{eqnarray}
with $x_{f}(1/2)=0$, we get
\begin{eqnarray}
x_{f}(t)={B\over A}(e^{{A\over 2}(2t-1)}-1). \label{exa3}
\end{eqnarray}
Then, substituting in Eq. (\ref{gensol24}), we obtain the general solution
\begin{eqnarray}
u(v,t)={1\over \sqrt{{2\pi\over
A}(1-e^{-2At})}}\int_{-\infty}^{+\infty}e^{-{\lbrace e^{-At}\lbrack
x-{B\over A}(e^{At}-1)\rbrack-y\rbrace^{2}\over {2\over
A}(1-e^{-2At})}}u(y,0)dy,\label{exa4}
\end{eqnarray}
where we have taken the origin of times at $t=0$.
 If we start from a
step function $u(x,0)=\eta(x)$, the above expression reduces to
\begin{eqnarray}
u(v,t)=\Phi\biggl \lbrack {x-{B\over A}(e^{At}-1)\over \sqrt{2\over
A}(e^{2At}-1)^{1/2}}\biggr \rbrack. \label{exa5}
\end{eqnarray}
On the other hand, for $A=0$, we get
\begin{eqnarray}
h(\tau)=0, \quad H(\tau)=0, \quad H_{1}(\tau)=1,\quad
H_2(\tau)=\tau-1,  \label{exa6}
\end{eqnarray}
\begin{eqnarray}
\chi^{2}(\tau)=2(\tau-1), \qquad x_{f}(t)={B\over 2}(2t-1).
\label{exa7}
\end{eqnarray}
The general solution is
\begin{eqnarray}
u(v,t)={1\over \sqrt{4\pi
t}}\int_{-\infty}^{+\infty}e^{-{(x-Bt-y)^{2}\over 4t}}u(y,0)dy,
\label{exa8}
\end{eqnarray}
and if we start from a step function $u(x,0)=\eta(x)$, we get
\begin{eqnarray}
u(v,t)=\Phi\biggl ({x-Bt\over 2t^{1/2}}\biggr ). \label{exa9}
\end{eqnarray}

We recall that the above profiles are the exact solutions of the
Fokker-Planck equation (\ref{gen1}) when the functions $U(v)$ and $D(v)$
satisfy the differential equation (\ref{relationDU1}). Let us examine some
particular cases. If $A=0$, Eq. (\ref{relationDU1}) can be
integrated at once leading to the relation
\begin{equation}
\label{CDU1} D(v)= B^{2} \left(\int_v^{+\infty} w^{d-1} e^{-U(w)}
dw\right)^2  {e^{2U(v)}\over v^{2(d-1)}}.
\end{equation}
For the wide class of potentials  $U(v)$ satisfying the
following condition:
\begin{equation}\label{CU1} (d-1){U'(v) \over v} - U''(v)  \ll  U'(v)^2 
\end{equation}
for large $v$, the behavior of the diffusion coefficient at infinity  is given by
\begin{equation}
D(v) \sim {C \over U'(v)^2}\, .
\end{equation}
In particular, when $U(v)=v^2/2$ the diffusion coefficient $D(v)$
behaves like $C/v^2$ for $v\rightarrow +\infty$ and when $U(v)=\gamma
v$ it tends to a constant.  Note that condition (\ref{CU1}) is not
only satisfied by any non constant polynomial, but also by potentials
$U(v)$ that dominate $\ln(v)$ for large velocities, as for instance
$U(v)=(\ln(v))^\alpha$, with $\alpha >1$. Note finally that for
$v\rightarrow 0$, the diffusion coefficient behaves like
$1/v^{2(d-1)}$ which is divergent except for $d=1$.

In the case of a quadratic potential $U(v)=v^{2}/2$, the integral in
Eq. (\ref{CDU1}) can be expressed in terms of the error
function. There is a simplification in $d=2$ leading to
$D(v)=1/v^{2}$. This is precisely the case that we have found in
Sec. \ref{sec_pw}. In that case, we have $x=v^{2}/2$ so that
Eq. (\ref{exa8}) can be explicitly written in terms of $v$.
Inversely, for $D(v)=1/v^{2}$ and $g=A$, we find from
Eq. (\ref{relationDU1}) that $U(v)=(A/8)v^4+(B/2) v^2+(d-2)\ln v$ so
for this type of potential the solution of (\ref{exa4}) applies with
$x=v^{2}/2$.  In the case of a linear potential $U(v)=\gamma v$, the
integral in Eq. (\ref{CDU1}) can be expressed in terms of the
incomplete Gamma function.  There is a simplification in $d=1$ leading
to $D(v)=1$. In that case, we have $x=v$ so that Eq. (\ref{exa8}) can
be explicitly written in terms of $v$. Inversely, for $D(v)=1$ and
$g=A$, we find from Eq. (\ref{relationDU1}) that $U(v)=(A/2)v^{2}+B
v+(d-1)\ln v$ so for this type of potential the solution of
(\ref{exa4}) applies with $x=v$.

For $d=1$, $D(v)=1$ and $U(v)={(A/2)}v^{2}+Bv$, we have $V(v)=Ax+B$
with $x=v$ and we are in the situation mentioned above. If we take
$f(v,0)=\delta(v-v_{0})$ and recall that $u(v,t)=f(v,t)/f_{e}(v)$,
Eq. (\ref{exa4}) gives after simplification
\begin{eqnarray}
f(v,t)={1\over
\sqrt{{2\pi\over A}(1-e^{-2At})}}e^{-{A\over
2(1-e^{-2At})}(v-v_{0}e^{-At}+{B\over A}(1-e^{-At}))^{2}}. \label{exa10}
\end{eqnarray}
In particular, for $B=0$ and $A=1$, we recover the well-known
solution \cite{risken}:
\begin{eqnarray}
f(v,t)={1\over
\sqrt{{2\pi}(1-e^{-2t})}}e^{-{(v-v_{0}e^{-t})^{2}\over
2(1-e^{-2t})}}. \label{exa11}
\end{eqnarray}
In that case, $\langle v\rangle(t) =v_0 e^{-t}$.
Alternatively, for $A=0$ and $B=1$, we get
\begin{eqnarray}
f(v,t)={1\over \sqrt{{4\pi}t}}e^{-{(v-v_{0}+t)^{2}\over 4t}}.
\label{exa12}
\end{eqnarray}
In that case, $\langle v\rangle(t) =v_0-t$. Note that there is no
normalizable stationary state for the Fokker-Planck equation
(\ref{gen1}) when $U(v)=B v$ so that $f(v,t)$ tends to zero for large
times and spreads so as to conserve the total mass. By contrast, when
$A\neq 0$, the distribution function relaxes towards $f_e(v)$ for
$t\rightarrow +\infty$. From Eq. (\ref{exa10}), we can obtain the time
evolution of the distribution function for any initial condition
$f_{0}(v)=f(v,0)$ by multiplying Eq. (\ref{exa10}) by $f_{0}(v_{0})$
and integrating over $v_{0}$.

Note finally that for $d>1$, the velocity $v=|{\bf v}|$ is restricted to
positive values so that the preceding results must be slightly
revised. The idea is to extend $f(v,t)$ by parity to negative values
of $v$. We replace $D(v)$, $U(v)$ and $v^{d-1}$ in Eq. (\ref{gen1}) by
$D(|v|)$, $U(|v|)$ and $|v|^{d-1}$. With this extension,
Eq. (\ref{gen1}) is invariant by the transformation $v\rightarrow
-v$. Therefore, if $f(v,t_{0})$ is even, $f(v,t)$ will remain even for
all times and its values for $v\ge 0$ correspond to those of the
distribution function that is solution of the original
Eq. (\ref{gen1}). Thus, we can apply the preceding results with almost
no modification. We note however that since $V(v)$ and
$x(v)=\int_{0}^{v}dw/\sqrt{D(w)}$ are odd, the case $V(v)=Ax+B$ is
only possible if $B=0$. By contrast, in $d=1$, we can consider cases
where $f(v,t)$, $D(v)$ and $U(v)$ have no special parity.

\section{Conclusion}
\label{sec_conclusion}

In this paper, we have developed a general formalism to characterize
the evolution of the distribution function tail for systems described
by a Fokker-Planck equation with a diffusion coefficient and a
friction force depending on the velocity. Our analytical results give
good agreement with the numerics even in cases where the validity of
our approach is marginal. When the diffusion coefficient decreases
algebraically with the velocity, the progression of the front is also
algebraic. When the diffusion coefficient decreases like an
exponential or like a Gaussian, the progression of the front is
logarithmic. The high velocity component of the distribution function
keeps the memory of the initial condition for a long time and is
slowly depleted. There are several applications of our formalism to,
e.g., stellar dynamics, plasma physics, vortex dynamics, the HMF
model, optical lattices etc. In future works, we shall study more
specifically the dynamics of point vortices  and
investigate the evolution of the front profile and the time evolution
of the correlation functions \cite{kinlemou}.

\newpage
\appendix

\section{General solution of Eq. (\ref{gen16})}
\label{sec_gensol}

In this Appendix, we provide the general solution of the PDE:
\begin{eqnarray}
2{\partial\phi\over\partial \tau}={\partial^{2}\phi\over\partial z^{2}}-h(\tau)z{\partial\phi\over\partial z},
\label{gensol1}
\end{eqnarray}
for an arbitrary initial condition $\phi_{1}(z)=\phi(z,1)$. Taking the Fourier transform of Eq. (\ref{gensol1}) with the conventions
\begin{eqnarray}
\phi(z)=\int \hat{\phi}(\xi)e^{i\xi z}d\xi,\qquad \hat{\phi}(\xi)=\int {\phi}(z)e^{-i\xi z}{dz\over 2\pi},\label{gensol2}
\end{eqnarray}
and using the relation
\begin{eqnarray}
z{\partial\phi\over\partial z}=\int z i \xi \hat{\phi}(\xi) e^{i\xi z}d\xi\qquad\qquad\nonumber\\
=\int \xi \hat{\phi}(\xi){\partial\over\partial \xi}(e^{i\xi z})d\xi=-\int {\partial\over\partial\xi}(\xi\hat{\phi}(\xi))e^{i\xi z}d\xi,\label{gensol3}
\end{eqnarray}
we get
\begin{eqnarray}
2{\partial\hat{\phi}\over\partial \tau}=(h(\tau)-\xi^{2})\hat{\phi}+h(\tau)\xi {\partial\hat{\phi}\over\partial\xi}.\label{gensol4}
\end{eqnarray}
We introduce the change of variables
\begin{eqnarray}
f(y,\tau)=\hat{\phi}(H_{1}(\tau)y,\tau), \qquad \xi=H_{1}(\tau)y\label{gensol5}
\end{eqnarray}
and choose the function $H_{1}(\tau)$ such that
\begin{eqnarray}
{H_{1}'(\tau)\over H_{1}(\tau)}=-{h(\tau)\over 2}.\label{gensol6}
\end{eqnarray}
Substituting Eq. (\ref{gensol5}) in Eq. (\ref{gensol4}), we find that $f(y,\tau)$ satisfies
\begin{eqnarray}
{\partial f\over\partial \tau}+{1\over 2}\lbrack H_{1}^{2}(\tau)y^{2}-h(\tau)\rbrack f=0.\label{gensol7}
\end{eqnarray}
Let $H(\tau)$ be the primitive of $h(\tau)$ such that
\begin{eqnarray}
H(\tau)=\int_{1}^{\tau}h(\tau')d\tau'.\label{gensol8}
\end{eqnarray}
Then, we choose $H_{1}$, solution of Eq. (\ref{gensol6}), such that
\begin{eqnarray}
H_{1}(\tau)=e^{-{H(\tau)\over 2}}.\label{gensol9}
\end{eqnarray}
By convention, $H(1)=0$ and $H_{1}(1)=1$. Equation (\ref{gensol7}) can be integrated leading to
\begin{eqnarray}
f(y,\tau)=f(y,1)e^{{H(\tau)\over 2}}e^{-{1\over 2}H_{2}(\tau)y^{2}},\label{gensol10}
\end{eqnarray}
where we have defined
\begin{eqnarray}
H_{2}(\tau)=\int_{1}^{\tau} H_{1}(\tau')^{2}d\tau'=\int_{1}^{\tau} e^{-H(\tau')}d\tau'.\label{gensol11}
\end{eqnarray}
Returning to original variables, we obtain
\begin{eqnarray}
\hat{\phi}(\xi,\tau)=\hat{\phi}_{1}\biggl ({\xi\over H_{1}(\tau)}\biggr )e^{{H(\tau)\over 2}}e^{-{1\over 2}{H_{2}(\tau)\over H_{1}^{2}(\tau)}\xi^{2}}.\label{gensol12}
\end{eqnarray}
We now observe that
\begin{eqnarray}
{H_{2}(\tau)\over H_{1}^{2}(\tau)}=\int_{1}^{\tau}
e^{H(\tau)-H(\tau')}d\tau'\equiv {1\over 2}\chi^{2}(\tau),\label{gensol13}
\end{eqnarray}
where $\chi(\tau)$ is the function introduced in Eq. (\ref{gen20}). Therefore
the general solution of Eq. (\ref{gensol1}) in Fourier space can be written
\begin{eqnarray}
\hat{\phi}(\xi,\tau)={1\over H_{1}(\tau)}\hat{\phi}_{1}\biggl
({\xi\over H_{1}(\tau)}\biggr )e^{-{1\over 4}\chi^{2}(\tau)\xi^{2}}.\label{gensol14}
\end{eqnarray}
Defining
\begin{eqnarray}
q(z)=\phi_{1}(H_{1}(\tau)z)\quad\leftrightarrow \quad \hat{q}(\xi)={1\over H_{1}(\tau)}\hat{\phi}_{1}\biggl({\xi\over H_{1}(\tau)}\biggr )
\end{eqnarray}
\begin{eqnarray}
g(z)=G(z/\chi(\tau))\quad\leftrightarrow \quad
\hat{g}(\xi)=\chi(\tau)\hat{G}(\chi(\tau)\xi) 
\end{eqnarray}
where
\begin{eqnarray}
G(z)=e^{-z^{2}}\quad\leftrightarrow \quad \hat{G}(\xi)={1\over 2\sqrt{\pi}}e^{-\xi^{2}/4}
\end{eqnarray}
we can rewrite Eq. (\ref{gensol14}) in the form
\begin{eqnarray}
\hat{\phi}(\xi,\tau)={2\sqrt{\pi}\over
\chi(\tau)}\hat{q}(\xi)\hat{g}(\xi). \label{gensol15}
\end{eqnarray}
Taking the inverse Fourier transform, we can express the solution of Eq. (\ref{gensol1})
as a convolution
\begin{eqnarray}
{\phi}(z,\tau)={2\sqrt{\pi}\over \chi(\tau)}\int
q(z-z')g(z'){dz'\over 2\pi}, \label{gensol16}
\end{eqnarray}
or, equivalently
\begin{eqnarray}
{\phi}(z,\tau)={1\over
\sqrt{\pi}}\int_{-\infty}^{+\infty}e^{-x^{2}}\phi_{1}(H_{1}(\tau)(z-\chi(\tau)x))dx.\label{gensol17}
\end{eqnarray}
By direct substitution, we can check that Eq. (\ref{gensol17}) is indeed solution of Eq. (\ref{gensol1}).

If $\phi_{1}(z)=\eta(z)$ is a step function with $\eta(z)=1$ for
$z<0$ and $\eta(z)=0$ for $z>0$, we immediately find that
\begin{eqnarray}
{\phi}(z,\tau)={1\over
\sqrt{\pi}}\int_{z/\chi(\tau)}^{+\infty}e^{-x^{2}}dx=\Phi\biggl
({z\over \chi(\tau)}\biggr ), \label{gensol18}
\end{eqnarray}
and we recover the result of Sec. \ref{sec_gen}. Then, the general
solution of Eq. (\ref{gensol1}) can be put in the form
\begin{eqnarray}
{\phi}(z,\tau)=\Phi\biggl ({z\over \chi(\tau)}\biggr )+{1\over
\sqrt{\pi}}\int_{-\infty}^{+\infty}e^{-x^{2}}(\phi_{1}-\eta)(H_{1}(\tau)(z-\chi(\tau)x))dx.\label{gensol19}
\end{eqnarray}
We introduce the new variable $z'=z/\chi(\tau)$ and consider the limit $t\rightarrow +\infty$. The integral depending on the initial condition is given by
\begin{eqnarray}
I={1\over
\sqrt{\pi}}\int_{-\infty}^{+\infty}e^{-x^{2}}(\phi_{1}-\eta)(\sqrt{2H_2(\tau)}(z'-x))dx.\label{gensol20}
\end{eqnarray}
We shall assume that $\phi_{1}(z)\rightarrow 1$ for $z\rightarrow
-\infty$ and $\phi_{1}(z)\rightarrow 0$ for $z\rightarrow
+\infty$. Then, if $H_2(\tau)\rightarrow +\infty$ for $\tau\rightarrow
+\infty$, the function $(\phi_{1}-\eta)(\sqrt{2H_2(\tau)}(z'-x))$ will be
very peaked around $x=z'$ and we can approximate the integral by 
\begin{eqnarray}
I\sim{e^{-z'^{2}}\over \sqrt{2\pi
H_2(\tau)}}\int_{-\infty}^{+\infty}(\phi_{1}-\eta)(x)dx \rightarrow
0. \label{gensol21}
\end{eqnarray}
Therefore, the condition that the solution of (\ref{gensol1}) tends
asymptotically to the function (\ref{gensol18}) for $\tau\rightarrow
+\infty$ is that $(\phi_{1}-\eta)(\pm \infty)=0$ and
\begin{eqnarray}
H_2(\tau)=\int_{1}^{\tau}e^{-H(\tau')}d\tau'\rightarrow +\infty
\label{gensol22}
\end{eqnarray}
for $\tau\rightarrow +\infty$. We can also write the general solution
(\ref{gensol17}) in the form
\begin{eqnarray}
{\phi}(z,\tau)={1\over\sqrt{2\pi
H_2(\tau)}}\int_{-\infty}^{+\infty}e^{-{(H_{1}(\tau)z-x)^{2}\over 2
H_2(\tau)}}\phi_{1}(x)dx.\label{gensol23} 
\end{eqnarray}
Returning to original variables, we get
\begin{eqnarray}
u(v,t)={1\over\sqrt{2\pi
H_2(2t)}}\int_{-\infty}^{+\infty}e^{-{(H_{1}(2t)(x-x_{f}(t))-y)^{2}\over
2 H_2(2t)}}u(y,1/2)dy. \label{gensol24} 
\end{eqnarray}
where $x=x(v)$ and we have taken by convention $x_{f}(t=1/2)=0$.

\section{Asymptotic behavior of Eq. (\ref{g2})}
\label{sec_ab}

Setting $z=(\gamma/2)y^{2}$, we can rewrite Eq. (\ref{g2}) in the
equivalent form
\begin{eqnarray}
x\sim {1\over \sqrt{2\gamma}}e^{{\gamma\over 2}v^{2}}\int_{0}^{{\gamma\over 2}v^{2}} {e^{z-{\gamma\over 2}v^{2}}\over \sqrt{z}}dz.
\end{eqnarray}
With the change of variables $y=-z+{\gamma\over 2}v^{2}$, we get
\begin{eqnarray}
x\sim {1\over \gamma v}e^{{\gamma\over 2}v^{2}}\int_{0}^{{\gamma\over 2}v^{2}} {e^{-y}\over \sqrt{1-{2y\over \gamma v^{2}}}}dy.
\end{eqnarray}
Then taking the limit $v\rightarrow +\infty$, we find that
\begin{eqnarray}
x\sim {1\over \gamma v}e^{{\gamma\over 2}v^{2}}\int_{0}^{+\infty}
{e^{-y}}dy\sim {1\over \gamma v}e^{{\gamma\over 2}v^{2}}.
\end{eqnarray}

\section{Criterion for the monotonicity of $f(v,t)$}
\label{sec_mono}

In this Appendix, we establish a criterion which guarantees the
monotonicity of $f(v,t)$ for all times, if the distribution function
is initially monotonic (decreasing). Let us first rewrite the
Fokker-Planck equation (\ref{gen1}) in the form
\begin{eqnarray}
{\partial f\over\partial t}=D{\partial^{2}f\over \partial v^{2}}+\left (D'+DU'+{d-1\over v}D\right ){\partial f\over\partial v}+\left\lbrack (DU')'+{d-1\over v}DU'\right\rbrack f.
\label{mono1}
\end{eqnarray}
Because of the positivity of the diffusion coefficient, a classical
comparison principle states that if $f(v,t)$ is initially positive, it
will remain positive for all times \cite{math}. The idea is now to
apply the same argument to $g=\partial f/\partial v$. Taking the
derivative of Eq. (\ref{mono1}) with respect to $v$, we get
\begin{eqnarray}
{\partial g\over\partial t}=D{\partial^{2}g\over \partial v^{2}}+\left (2D'+DU'+{d-1\over v}D\right ){\partial g\over\partial v}+\left\lbrack D''+2(DU')'+{d-1\over v}DU'+(d-1)\left ({D\over v}\right )'\right\rbrack g\nonumber\\
+\left\lbrack (DU')''+(d-1)\left ({DU'\over v}\right )'\right\rbrack f.\qquad
\label{mono2}
\end{eqnarray}
If
\begin{eqnarray}
(DU')''+(d-1)\left ({DU'\over v}\right )'\le 0,
\label{mono3}
\end{eqnarray}
then
\begin{eqnarray}
{\partial g\over\partial t}\le D{\partial^{2}g\over \partial v^{2}}+\left (2D'+DU'+{d-1\over v}D\right ){\partial g\over\partial v}+\left\lbrack D''+2(DU')'+{d-1\over v}DU'+(d-1)\left ({D\over v}\right )'\right\rbrack g.\nonumber\\
\label{mono4}
\end{eqnarray}
Again we use a comparison principle and deduce from this inequality that if $g(v,t)$ is initially
negative, it will remain negative for all times. Therefore, if
$f(v,t)$ is initially decreasing, it will remain decreasing for all
times if the criterion (\ref{mono3}) is satisfied. This criterion is
just a {\it sufficient} condition of monotonicity.

Let us give some particular examples of application. In the case of a
quadratic potential $U(v)=v^{2}/2$, the criterion (\ref{mono3})
becomes
\begin{eqnarray}
(Dv)''+(d-1)D'\le 0.
\label{mono5}
\end{eqnarray}
For a diffusion coefficient of the form $D=v^{-\alpha}$, we find the
condition $\alpha\le d$. In particular, for the case $\alpha=d=3$, the
criterion is satisfied. We can check that the criterion (\ref{mono5})
is also satisfied for all $v$ for the diffusion coefficient
(\ref{ex6}-a) corresponding to the Coulombian (or Newtonian) potential
of interaction.  Alternatively, for a diffusion coefficient of the
form $D=e^{-\gamma v^{2}}$, we find that the criterion (\ref{mono5})
is not satisfied for large values of $v$ (more precisely
$v>(d+\sqrt{d^{2}+8\gamma})/4\gamma$). Therefore, the profile $f(v,t)$
can be non-monotonic even if the initial condition is monotonic, as in
the case of Fig. \ref{shortt}.

\section{General solution of Eq. (\ref{relationDU1})}
\label{sec_gdu}

In this Appendix, we show that Eq. (\ref{relationDU1}) can be solved
explicitly. First of all, we note that it can be written
\begin{eqnarray}
\left (
U'(v)-{d-1\over
  v}\right )R-R'(v)=Ax+B.
\label{fin1}
\end{eqnarray}
Since $R=dv/dx$, we obtain
\begin{eqnarray}
\left (U'(v)-{d-1\over
  v}-{R'(v)\over R(v)}\right )dv=(Ax+B)dx
\label{fin2}
\end{eqnarray}
which leads to
\begin{eqnarray}
e^{-U(v)}v^{d-1}R(v)=Ke^{-({A\over 2}x^{2}+Bx)},
\label{fin3}
\end{eqnarray}
where $K$ is a constant. Equation (\ref{fin3}) can again be
integrated into
\begin{eqnarray}
\int_{v}^{+\infty}e^{-U(w)}w^{d-1}dw=K\int_{x(v)}^{+\infty}e^{-({A\over 2}y^{2}+By)}dy.
\label{fin4}
\end{eqnarray}
For $A=0$, we obtain 
\begin{eqnarray}
\int_{v}^{+\infty}e^{-U(w)}w^{d-1}dw={K\over B}e^{-Bx}.
\label{fin5}
\end{eqnarray}
Substituting this relation in Eq. (\ref{fin3}), we recover
the result of Eq. (\ref{CDU1}). On the other hand, for $A\neq 0$, we
get
\begin{eqnarray}
\int_{v}^{+\infty}e^{-U(w)}w^{d-1}dw=K\sqrt{2\pi\over A}e^{B^{2}\over 2A}\Phi\biggl\lbrack \sqrt{A\over 2}\biggl ({B\over A}+x\biggr )\biggr\rbrack,
\label{fin6}
\end{eqnarray}
where $\Phi(x)$ is defined in Eq. (\ref{gen19}). Substituting the
foregoing relation in Eq.  (\ref{fin3}), we find that
\begin{eqnarray}
D(v)={e^{2U(v)}\over v^{2(d-1)}}F\left\lbrack \int_{v}^{+\infty} C e^{-U(w)}w^{d-1}dw\right\rbrack,
\label{fin7}
\end{eqnarray}
where 
\begin{eqnarray}
F(x)={\rm exp}\left\lbrace -2(\Phi^{-1}(x))^{2}\right\rbrace,
\label{fin9}
\end{eqnarray}
and $C$ is a constant.

\end{document}